\documentclass[journal,comsoc,twoside]{IEEEtran}

\usepackage{graphicx}
\usepackage{comment}
\usepackage[table]{xcolor}
\usepackage{balance}
\usepackage{caption}
\usepackage{subcaption}
\usepackage{url}

\usepackage{mathtools}  
\usepackage{tabulary}
\usepackage{booktabs}

\usepackage{amssymb}
\usepackage{amsfonts}
\usepackage{amsmath}
\usepackage{amsthm}

\newtheorem{theorem}{Theorem}

\usepackage[ruled,vlined,linesnumbered,longend]{algorithm2e}

\usepackage{graphicx}
\usepackage{caption}
\usepackage{subcaption}
\usepackage{multirow}

\begin{document}

\title{Distributed Data Access in Industrial Edge Networks}

\author{Theofanis P. Raptis, Andrea Passarella, Marco Conti%
\thanks{The authors are with the Institute of Informatics and Telematics, National Research Council, Pisa, Italy.
}%
\thanks{This work was funded by the European Commission through the FoF-RIA Project AUTOWARE: Wireless Autonomous, Reliable and Resilient Production Operation Architecture for Cognitive Manufacturing (No. 723909).}%
}


\maketitle

\begin{abstract}
Wireless edge networks in smart industrial environments increasingly operate using advanced sensors and autonomous machines interacting with each other and generating huge amounts of data. Those huge amounts of data are bound to make data management (e.g., for processing, storing, computing) a big challenge. Current data management approaches, relying primarily on centralized data storage, might not be able to cope with the scalability and real time requirements of Industry 4.0 environments, while distributed solutions are increasingly being explored. In this paper, we introduce the problem of distributed data access in multi-hop wireless industrial edge deployments, whereby a set of consumer nodes needs to access data stored in a set of data cache nodes, satisfying the industrial data access delay requirements and at the same time maximizing the network lifetime. We prove that the introduced problem is computationally intractable and, after formulating the objective function, we design a two-step algorithm in order to address it. We use an open testbed with real devices for conducting an experimental investigation on the performance of the algorithm. Then, we provide two online improvements, so that the data distribution can dynamically change before the first node in the network runs out of energy. We compare the performance of the methods via simulations for different numbers of network nodes and data consumers, and we show significant lifetime prolongation and increased energy efficiency when employing the method which is using only decentralized low-power wireless communication instead of the method which is using also centralized local area wireless communication.
\end{abstract}

\begin{IEEEkeywords}
Industry 4.0, Data Caching, Internet of Things, Energy Efficiency, Wireless Networks
\end{IEEEkeywords}

\section{Introduction and Motivation}

\IEEEPARstart{I}{n} the frame of cyber-physical convergence \cite{CONTI20122}, industrial equipment will be more and more saturated with embedded devices generating large amounts of data as the basis for real-time control and operations in factory environments. Recent initiatives such as the ``Industry 4.0''  in Europe, the ``Industrial Internet'' in United States, the ``Manufacturing White Book of Year 2014'' in Japan, and the ``Made in China 2025'' in China, have set the foundations for the deployment of \emph{large scale wireless edge networks} as a technical enabler for a variety of industrial use cases \cite{8012376}. Wireless edge networks in smart industrial environments operate using advanced sensors and information technologies; thus, large amounts of data are generated and collected, requiring big data processing technology to build an integrated environment in which the automation process can be represented transparently and controlled and managed in a more efficient way \cite{8085101}. 

More and more, in industrial environments, Internet of Things (IoT) devices are not seen as ``dumb things'' generating individually a few bytes, but as small multimedia devices, possibly generating data of significant size (e.g., high resolution pictures or video), still operated via batteries to make them more flexible and cheap to assemble, install and manage \cite{networld2020}. Standardized protocols, such as WirelessHART and ISA100 Wireless, provide the basis for energy-efficient mesh/multi-hop communications, while reducing fading and interference impairments \cite{7851047}. In this setting, \emph{how to process, store, and manage the data efficiently, while at the same time satisfy the industrial requirements}, is a great challenge. 

The role of data in Industry 4.0 wireless networks is becoming more and more important. Industrial applications generate more and more data, both at each individual node (e.g., multimedia IoT devices) and at the entire factory level, due to the increasing number of devices embedded in industrial physical components. Moreover, several key Industry 4.0 automation applications are enabled by efficient flowing of data across nodes. Therefore, a very important perspective is seeing industrial networks as data-centric, and including efficient data management as a first class primitive in such networking environments.

Traditionally, cloud computing provides us with an opportunity to solve the related data-centric problems centrally and seamlessly. In many cases, hierarchical clouds or fog computing tools are also employed for additional functionality \cite{8259028}. Although there has been considerable work on caching of and accessing industrial edge data on the cloud \cite{8259028}, the burden of transferring the data back and forth between local and remote locations has already been demonstrated as excessive for some use cases and prohibiting for others \cite{8472907}. A representative example is the 5GPPP classification of industrial use cases \cite{5gppp} in five families, each of them representing a different subset of communication requirements in terms of delay, reliability, availability, throughput, etc. In this classification, most of the time critical use cases, such as motion control, control-to-control and massive wireless sensor networks, \emph{necessitate end-to-end data delivery delays ranging from 1 to 100 ms}. Even some cases of employment of pure edge data distribution methods can effectively lead to sub-optimal performance, such as the case of ultra low delay applications \cite{Raptis_2018}. Last but not least, with the increasing number of involved \emph{battery-powered devices}, industrial edge deployments may consume substantial amounts of energy; more than needed if local, distributed and low-power computations were used instead. This is due to the fact that low-power wireless communication can be way more costly than local area wireless.

Because of that, more and more decentralized solutions to data management are explored, whereby centralized controllers, cache nodes and also individual, resource-constrained sensor motes are exploited to cache data and serve them to consumers on demand. To support this, novel data management layers need to be engineered over the device and networking planes of the industrial deployments \cite{Lucas_Esta__2018,7785890}. Those layers operate independently from and complement the routing process, targeting at distributing the data in the network in a decentralized manner, while at the same time respecting the strict Industry 4.0 requirements. A fundamental operation of a data management layer on industrial edge networks is the \emph{efficient caching of industrial data in intermediary cache nodes which are located between the data source nodes and the data consumer nodes}. Typically, the cache nodes are more powerful than the resource-constrained edge sensor motes in terms of processing, communication and energetic autonomy capabilities. 

The combination of three elements, namely the industrial devices (edge devices and cache nodes), the industrial edge communication (wireless, low-power, multi-hop), and the industrial requirements (delay constraints, energy restrictions, and data pieces in need to be accessed by consumers), uncovers a new computational problem: \emph{Find the appropriate cache node for each data piece so that it can be effectively stored for future access by the interested consumers, while respecting the data access delay threshold, and maximizing the network lifetime.}

The fact that renders this specific, realistic problem particularly interesting for investigation is that, although similar problems have been addressed in the general ICT literature (graph theory, operations research, IoT), to the best of our knowledge, the exact problem and its formulation, which is naturally emerging from the combination of those three elements in industrial environments, \emph{has not been investigated in the past}. Specifically, from the theoretic point of view, the problem introduced in this paper presents remote similarities with multi-commodity flow problems \cite{4567876}. However, the presence of cache nodes differentiates significantly the network structure and the formulation properties. On a more practical basis, multi-source, multi-destination approaches have also been studied in the wireless sensor networking and internet of things domains \cite{5677539}. The problems investigated in those fields present even more similarities than the theoretic ones, as they usually involve lifetime maximization \cite{6687258} or delay considerations \cite{7950207}. Similar to the theoretic case however, the necessity for cache nodes alters fundamentally the network model and the combination of the specific objective function and problem constraints. As far as the pure industrial networking literature is concerned, the problem presented in this paper is also introduced for the first time. Although most recent approaches, bear some common elements individually (lifetime maximization in \cite{8325491}, delay constraints in \cite{6812138}, cache nodes in \cite{Raptis_2018}), they focus on different network models and problems. Deeper insights on the specific models and problems can be found in  \cite{8764545}, a truly extensive survey for future reference and open challenges.

In this paper we contribute to fill this gap. Specifically, extending our preliminary work presented in \cite{8390794}, we present a distributed, adaptive data access scheme for guaranteeing real-time delay requirements while effectively prolonging network lifetime in industrial edge wireless networks. The remainder of the paper is organized as follows. In section \ref{sec::basics}, we describe the technological background of the industrial edge networks and we sketch the network model abstraction that we use in this paper. In section \ref{sec::dcda}, we prove that the distributed data caching problem for industrial edge networks is computationally intractable, and we formulate the objective function on the maximum lifetime, which in our case is the time elapsed until the first node in the network runs out of energy. In section \ref{sec::algorithm}, we design a two-step algorithmic method for effectively prolonging the network lifetime, while respecting the industrial data access delay constraints. We use FIT IoT-LAB open testbed \cite{7389098} for conducting an experimental investigation. In order to demonstrate the efficiency of our method, and due to the fact that, to the best of our knowledge, there are no related solutions in the literature for the exact problem formulation, we compare its performance to the performance of an optimal solution of a relaxed formulation (and thus more difficult version) of the problem. Results show a very good accuracy of our solutions in approximating the optimal solution even of the relaxed problem. In section \ref{sec::impro}, we go one step beyond the strict, constant data distribution schedule and we provide two online improvements, so that the data distribution can dynamically change before the first node in the network dies: one centralized method which uses local area wireless communication to renew the data distribution schedules according to the current network energy map and one distributed method which periodically rotates the available data distribution paths in a proportionally fair manner. We then compare the performance of all three methods via simulations for different numbers of network nodes and data consumers, by using three metrics; network lifetime, energy consumption rate and total variation distance. Finally, in section \ref{sec::conc}, we conclude the paper.

\section{Industrial Edge Background} \label{sec::basics}

In this section, we describe the technological background of the industrial edge networks and we sketch the network model abstraction that we use in this paper. The notation used throughout the paper is summarized in Table \ref{tab::notation}.

\begin{table}[t!]\caption{Notation used in the paper}
\centering 
\begin{tabular}{c l p{6cm} }
\toprule
\multicolumn{3}{c}{}\\
\multicolumn{3}{c}{\underline{Basics}}\\
\multicolumn{3}{c}{}\\
$V$ & $\triangleq$ & set of network nodes $u$\\
$P$ & $\triangleq$ & set of cache nodes $p$, with $P \subset V$, $|P| \ll |V-P|$, and $E_p \gg E_u, \forall u \in V, p \in P$\\
$(u,v)$ & $\triangleq$ & wireless link between $u,v \in V$\\
$\delta(u,v)$ & $\triangleq$ & Euclidean distance between $u,v \in V$\\
$N_u$ & $\triangleq$ & neighborhood of $u$\\
$\gamma$ & $\triangleq$ & neighborhood adjustment parameter\\
$E_u$ & $\triangleq$ & energy supply of $u$\\
$D$ & $\triangleq$ & set of data pieces $d$\\
$\epsilon_{uv}$ & $\triangleq$ & data piece propagation on $(u,v)$ energy cost\\
$\tau$ & $\triangleq$ & time cycle\\
$s_d$ & $\triangleq$ & source node of $d$\\
$c_d$ & $\triangleq$ & consumer node of $d$\\
$r^g_d$ & $\triangleq$ & data generation rate of $d$\\
$r^c_d$ & $\triangleq$ & data consumption rate of $d$\\
$l_{uv}$ & $\triangleq$ & one hop data propagation delay\\
$L_{uv}$ & $\triangleq$ & end-to-end data propagation delay\\
$L_{\text{max}}$ & $\triangleq$ & end-to-end data access delay threshold\\
\multicolumn{3}{c}{}\\
\multicolumn{3}{c}{\underline{Decision variables}}\\
\multicolumn{3}{c}{}\\
$x^{c_d}_{uv}$ & $=$ & \(\begin{cases}
1,  & \text{if $(u,v)$ is active in path between $c_d$ and $p$} \\
0,  & \text{otherwise} \end{cases}\)\\
\multicolumn{3}{c}{}\\
$x^{s_d}_{uv}$ & $=$ & \(\begin{cases}
1,  & \text{if $(u,v)$ is active in path between $s_d$ and $p$} \\
0,  & \text{otherwise} \end{cases}\)\\
\multicolumn{3}{c}{}\\
\multicolumn{3}{c}{\underline{Objective function}}\\
\multicolumn{3}{c}{}\\
$a_{uv}$ & $\triangleq$ & $\sum_{i = 1}^{\nu} (r^g_i x^{s_i}_{uv} + r^c_i x^{c_i}_{uv})$, aggregate data rate of $(u,v)$\\
$\mathbf{x}$ & $\triangleq$ & $[a_{uv}]$, data rate vector of $u$\\
$T_u(\mathbf{x})$ & $\triangleq$ & $\frac{E_u}{\sum_{v \in N_u} \epsilon_{uv} a_{uv}}$, lifetime of $u$\\
$T(\mathbf{x})$ & $\triangleq$ & $\min_{u \in V} \left\{ T_u(\mathbf{x}) \mid \sum_{v \in N_u} (x^{s_d}_{uv} + x^{c_d}_{uv}) > 0\right\}$, network lifetime\\
\bottomrule
\end{tabular}
\label{tab::notation}
\end{table}

\begin{figure}[t!]
    \centering
         \includegraphics[width=\columnwidth]{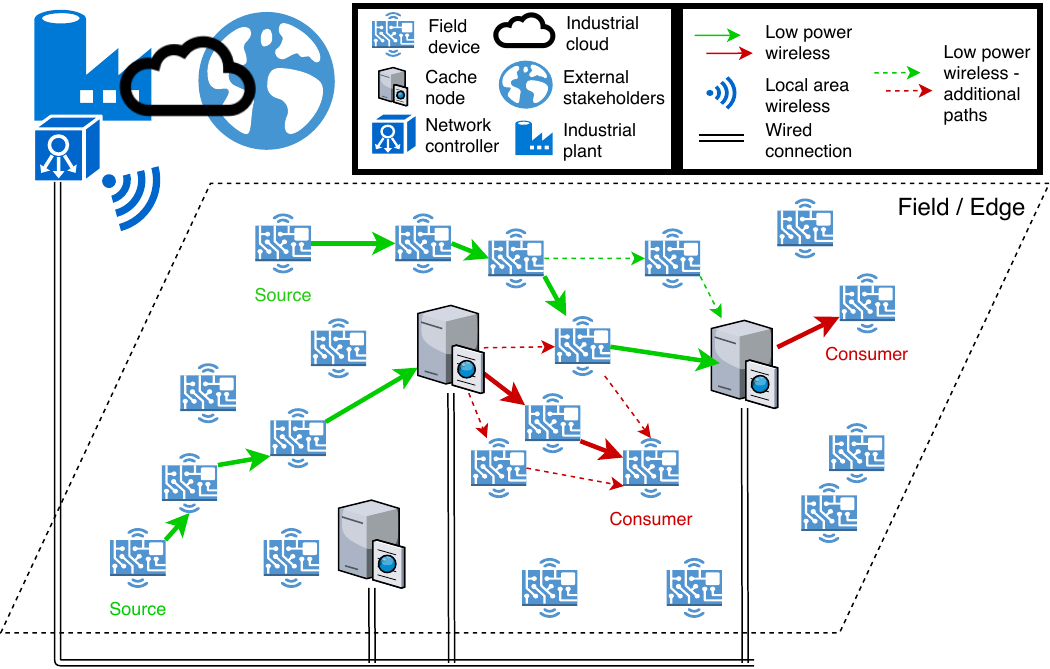}
        \caption{Industrial edge deployment.}
        \label{fig::arch}
\end{figure}

\textbf{Industrial devices}: We consider industrial edge deployments as shown in Fig. \ref{fig::arch}. The industrial plant is the center of operations and is connected to the industrial field on the one end and to the external stakeholders on the other end. The industrial plant features a network controller and the field contains the industrial network. The industrial network typically features two types of devices, as shown in Fig. \ref{fig::arch}: a large number of resource-constrained sensor and actuator nodes (which are equipped with limited processing, storing and communication abilities) and a limited number of more powerful cache nodes (which can be devices like Raspberry Pi or TTTech's MFN 100, or even devices such as Dell or Intel IoT gateways or Siemens PLCs). Such types of industrial networks are particularly related to smart factories \cite{8333734} and industrial workshop deployments \cite{8291116}.

\textbf{Industrial Communications}: As shown in Fig. \ref{fig::arch}, various communication and networking technologies co-exist in the industrial edge deployment. Typically, the communication between the industrial plant and the external stakeholders is performed via an industrial cloud system and relevant services (the investigation of data distribution aspects on this part of the communications is outside the scope of this paper). The communication between the industrial plant and the field network is performed through the central network controller. This part of the communication is implemented using two kinds of technologies, wired and wireless. The wired part is facilitating the communication between the central network controller with the cache nodes (e.g., TSN, HART). The wireless part is provided by two technologies: local area radio for the inter-field communication between the edge nodes and the controller (e.g., IEEE 802.11) and low power radio for the intra-field communication among the edge nodes (e.g., IEEE 802.15.4).

\textbf{The network}: We are particularly interested in the industrial edge network of the field devices and cache nodes, as well as in the associated data which are generated, cached and accessed. The network can be abstracted by a graph $G=(V,E)$ where every field device (or node) $u \in V$ has a limited amount of energy supply $E_u$. The more powerful (in terms of computation, storage and energy supplies) cache nodes are represented by a set $P$, with $P \subset V$, $|P| \ll |V-P|$, and $E_p \gg E_u, \forall u \in V, p \in P$. Cache nodes $p \in P$ act as local memories in the network and store data originated from the sources, for access from the consumers when needed. This relieves the field devices from the burden of storing data they generate (which might require excessive local storage), and helps meeting delay constraints. A node $u \in V$ can propagate data using suitable industrial networking standards (e.g., ISA100 Wireless, WirelessHART) to a set of nodes which lie in its neighborhood $N_u$. We assume that transmissions are sufficiently reliable and do not cause excessive interference. One way of capturing this assumption is via the formulation of $N_u$. $N_u$ contains the nodes $v \in V$ for which it holds that $\gamma \cdot \rho_u \geq \delta(u,v)$, where $\rho_u$ is the transmission range of node $u$ (defined by the output power of the antenna), $\delta(u,v)$ is the Euclidean distance between $u$ and $v$, and $\gamma$ is a neighborhood adjustment parameter, calibrated by the network operator, with $0 < \gamma \leq 1$. Parameter $\gamma$ is particularly useful in indoor industrial environments where high amounts of interference exist \cite{1435743}, and the nodes might need to transmit messages only to other nodes which are located nearby. The sets $N_u$ are thus defining the set of edges $E$ of the graph $G$. Each one-hop data propagation from $u$ to $v$ requires an amount of $\epsilon_{uv}$ of energy dissipated by $u$ so as to transmit one data piece (for example, a few bytes) to $v$. A schedule of multiple one-hop data propagations over different nodes in the network is resulting in multi-hop data routing, according to the underlying routing protocol. Finally, we also assume the existence of a wireless local area network, whereby all devices can communicate wirelessly with the controller. The wireless local area communication, however, is much more power hungry and thus more costly, as for example, low-power edge wireless links are typically operating at $-25$ dBm whereas local area wireless links at $15$ dBm.

\textbf{The data}: The generated data in the network are modeled as a set of data pieces $D$. Each data piece $d \in D$ is defined as $d = (s_d, c_d, r^g_d, r^c_d)$, where $s_d \in V$ is the source of data piece $d$, $c_d \in V$ is the consumer of data piece $d$, and $r^g_d$, $r^c_d$ are the data generation and consumption rates of $d$ respectively. If the same data of a source, e.g., $s_1$, is requested by more than one consumers, e.g., $c_1$ and $c_2$, we have two distinct data pieces, $D_1 = (s_1, c_1, r^g_1, r^c_1)$ and $D_2 = (s_2=s_1, c_2, r^g_1, r^c_2)$.  Applications that fit this data model, e.g., continuous monitoring, involve data coming from multiple devices and are needed by a single device after asynchronous consumption requests, without necessarily a strictly defined pattern of data generation and requests (which is why we assume a rate of generation and consumption). Without loss of generality, we divide time in time cycles $\tau$ and we assume that the data may be generated at each source $s_d$ and consumed at each consumer $c_d$ (according to rates $r^g_d, r^c_d$) at the beginning of each $\tau$. Naturally, being asynchronous, the data generated need to be cached temporarily for future requests by consumers. 

\textbf{The delay}. Each one-hop data propagation from $u$ to $v$ results in a delay $l_{uv}$. We denote as $L_{uv}$ the end-to-end delay of the multi-hop data propagation from $u$ to $v$, where $L_{uv} = l_{uw} + ... + l_{zv}$, where $w \in N_u$ and $z \in N_v$. A critical aspect in the industrial operation process is the timely data access by the consumers. The industrial requirements impose a maximum data access delay constraint which introduces an end-to-end data access delay threshold $L_{\text{max}}$ for each consumer. We denote as $L_{p_dc_d}$ the data access delay of the consumer $c_d$ to the data piece $d$, with $L_{p_dc_d} = l_{p_dw} + ... + l_{zc_d}$, where $p_d$ is the cache node which holds $d$ for delivery to $c_d$. In order to meet the industrial requirements, we need to guarantee that, when a new request for a data piece is generated at a consumer, the data piece can be sent from the cache and reach the consumer within that maximum delay, and consequently, the following constraint must be met:
\begin{equation}
L_{p_dc_d} \leq L_{\text{max}}, \forall c_d \in V \label{eq::constr}
\end{equation}
Note that efficient data access delay is a particularly important requirement of future network-based communications for many targeted industrial applications in the concept of the Industry 4.0. Both the \emph{WG1 of Plattform Industrie 4.0} (reference architectures, standards and norms) \cite{reqs} and the \emph{Expert Committee 7.2 of ITG} (radio systems) \cite{itg} set the delay requirements for condition monitoring applications to around $100$ ms. An additional delay threshold could be imposed on the end-to-end data delivery from the source nodes to the corresponding cache nodes, or in other words the data caching delay ($L_{s_dp_d}$). The data caching delay, although not affecting the access delay of the consumers to the cached data, can affect the freshness of the cached data. As the freshness of the cached data is beyond the scope of this study, for simplicity we assume that $L_{s_dp_d}$ can remain unbounded. However, the modeling presented in this paper can be generalized so as to include also constraints on the data caching delay.

Following this modeling, the data caching process is comprised of two intertwined open issues: find the appropriate cache node $p_d$ for each $d$ so that it can be effectively stored for future use by $c_d$ while respecting the data access delay threshold $L_{\text{max}}$, and find an efficient data distribution method so that the network maximizes its lifetime.

\section{The Distributed Data Access Problem (\texttt{DDA})} \label{sec::dcda}

In this section, we prove that the distributed data caching problem for industrial edge networks is $\mathcal{N}\mathcal{P}$-complete (while it is true that similar problems in communication and network optimization can be intractable, we did not find any already existing exact problem formulation overlapping with the formulation that we provide for the problem presented in this paper; for this reason, we analytically prove its intractability). Suppose that we are given a network $G = (V,E)$, a set of cache nodes $P$ deployed in the network, the energy supplies $E_u$ and energy consumption costs $\epsilon_{uv}$ for every $u,v \in V$, and a fixed number of $\nu$ data pieces. The Distributed Data Access (\texttt{DDA}) problem is to find a feasible multi-hop data distribution schedule such that at least one $p \in P$ stores the data for every consumer $c_d \in V$, the delay constraint $L_{p_dc_d} \leq L_{\text{max}}$ is met for every $c_d \in V$, and the network lifetime is maximized. We consider as network lifetime the time point when the first field node in the network runs out of the sufficient energy supplies needed for proper operation, which is a typical measure of network lifetime in the networking literature \cite{ANASTASI2009537}. 

\subsection{Computational intractability of the problem}

In order to demonstrate that \texttt{DDA} is a difficult problem to solve, we formulate it as a decision problem, i.e., one where we check the existence of a feasible allocation (which can be posed as a yes-no question of the input values). We then show that even the decision version is computationally intractable; this result gives us leverage on designing efficient algorithms for the computation version of the problem. The decision version of the problem introduces parameter $T$, which is a time value, and poses the following question: can we solve \texttt{DDA} and no node in the network runs out of energy before time $T$? Note that $T$ is used for the proof on $\mathcal{N}\mathcal{P}$-completeness, and will be omitted in the formulation of the computation version, as the problem will be turned into a maximization of the time until the first node dies. We show that the general version of the \texttt{DDA} is $\mathcal{N}\mathcal{P}$-complete.

\begin{theorem} \label{theo::theo}
The decision version of the \texttt{DDA} problem is $\mathcal{N}\mathcal{P}$-complete.
\end{theorem}

\begin{figure}[t!]
    \centering
    \begin{subfigure}[b]{0.49\columnwidth}
        \includegraphics[width=\textwidth]{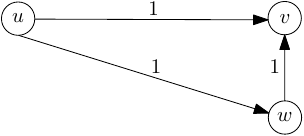}
        \caption{\texttt{D2CIF} graph}
        \label{fig::break1}
    \end{subfigure}
    \begin{subfigure}[b]{0.49\columnwidth}
        \includegraphics[width=\textwidth]{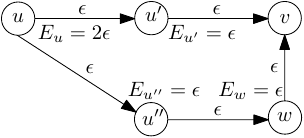}
        \caption{\texttt{DDA} node insertion.}
        \label{fig::break2}
    \end{subfigure}
    \caption{Toy example of a graph transformation in \texttt{DDA}.}\label{fig::break}
\end{figure}

\proof
We first note that, given a certain data distribution schedule, we can verify whether this schedule is sufficient so that no node dies, i.e., no node $u$ spends exactly $E_u$ energy for propagating data before time $T$. In particular, this can be done in $\mathcal{O}(\nu V)$ time, by summing the energy spent on every node in the network for every data piece generated and distributed until time $T$. Therefore, we have that \texttt{DDA} $\in \mathcal{N}\mathcal{P}$. For the hardness part we use the Directed Two Commodity Integral Flow problem \cite[p.~216]{Garey}, \texttt{D2CIF} in short. Let $G(V,A), s_1, s_2, t_1, t_2, c(a) = 1, \forall a \in A, R_1, R_2$ be the input of \texttt{D2CIF}. We now transform this into an input for \texttt{DDA} as follows: We set the energies needed for the transmission of a data piece $\epsilon_{uv} = \epsilon \cdot c(u,v) \overset{c(a) = 1, \forall a \in A}{=} \epsilon$, the delays required for the one-hop data propagations of a data piece $l_{uv} = l, \forall u,v \in V$ and the initial energy supplies of the nodes in the network $E_u = \sum_{v \in N_u} \epsilon_{uv}, \forall u,v \in V$. Then, we modify $G$ in order to take into account the edge capacities of \texttt{D2CIF} as follows: For every $u \in V$ with $E_u > \epsilon$, we ``break'' the edge $(u,v)$ and we insert an additional node $u'$ with $E_{u' }= \epsilon$, as shown in Fig.~\ref{fig::break}, thus replacing edge $(u,v)$ with two new edges, $(u,u')$ and $(u',v)$, with energies needed for transmission of a data piece $\epsilon_{uu'} = \epsilon_{u'v} = \epsilon$. We set the number of data pieces $ \nu = d_1 + d_2$ with $d_1 = R_1$ data pieces $(s_1, c_1, 1, 1)$ and $d_2 = R_2$ data pieces $(s_2, c_2, 1, 1)$. We set $c_i = p_i = t_i$ and $L_{\text{max}} < \min_{u \in N_{c_i}}{l_{c_iu}}, \forall i \in \{1,2\}$. Finally, assuming that the data are generated at the beginning of each time cycle $\tau$,  we set $T = \tau$. In order to reassure that the data pieces of cycle $\tau$ have been delivered before the start of cycle $\tau +1$, $\tau$ has to be sufficiently long. For this reason we set $\tau = l \cdot |E|$. Notice that an answer to this instance of the \texttt{DDA} problem would provide an answer also to \texttt{D2CIF}, which means that \texttt{D2CIF} $\leq_m$ \texttt{DDA}. This completes
 the proof.
\endproof

\subsection{The objective function on the maximum lifetime} \label{sec::obj}

In order to formalize the computation version of the problem, we need to define the target objective function, which is the maximization of the network lifetime. In order to construct the objective function, we build on top of existing typical sensor networking formulations like the ones presented in \cite{1331424} and \cite{4573259}. We introduce the variables, $x^{c_d}_{uv}$ and $x^{s_d}_{uv}$ which hold the necessary information regarding the distribution of the data pieces across the edges of the graph $G$. More specifically, $x^{c_d}_{uv} =1$ when edge $(u,v)$ is activated for data piece $d$ along an edge of a path between consumer $c_d$ and the corresponding cache node $p$. On the contrary, $x^{c_d}_{uv}=0$ when edge $(u,v)$ is inactive for the transmission of data piece $d$. Similarly, $x^{s_d}_{uv} =1$ when edge $(u,v)$ is activated for data piece $d$ along an edge of a path between source $s_d$ and the corresponding cache node $p$ and $x^{s_d}_{uv}=0$ when edge $(u,v)$ is inactive. We denote as $a_{uv} = \sum_{i = 1}^{\nu} (r^g_i x^{s_i}_{uv} + r^c_i x^{c_i}_{uv})$ the aggregate data rate of $(u,v)$ for all $\nu$ data pieces. Stacking all $a_{uv}$ together, we get $\mathbf{x} = [a_{uv}]$, the data rate vector of node $u$ for every $v \in N_u$. Following this formulation, and recalling that every one-hop data piece propagation on $(u,v)$ costs $\epsilon_{uv}$, the lifetime of node $u \in V$ can be defined as
\begin{equation} 
T_u(\mathbf{x}) = \frac{E_u}{\sum_{v \in N_u} \epsilon_{uv} a_{uv}}
\end{equation}
and the network lifetime which is also the objective function of the problem as
\begin{equation}
T(\mathbf{x}) = \min_{u \in V} \left\{ T_u(\mathbf{x}) \mid \sum_{v \in N_u} (x^{s_d}_{uv} + x^{c_d}_{uv}) > 0\right\}. \label{eq::obj}
\end{equation}

\section{A Centralized Algorithm for Solving the \texttt{DDA} Problem} \label{sec::algorithm}

Due to the $\mathcal{N}\mathcal{P}$-completeness and the inherent, proven intractability of the problem, optimal solutions which run in polynomial time are not attainable. For this reason, in this paper we followed the approach of designing a highly efficient method which sufficiently approximates the theoretically optimal performance, and of realistically validating its performance through experiments with field devices and simulations. Specifically, we design an algorithmic method for solving the computation version of the \texttt{DDA} problem; in other words, a method which approximates the network lifetime as defined by equation \ref{eq::obj}, while respecting the data access delay constraint of equation \ref{eq::constr}. The method is divided in two steps. In the first step, for each data piece, we compute a set of data distribution paths in the network which satisfy the delay constraint and in the second step we configure the variables $x^{c_d}_{uv}$ and $x^{s_d}_{uv}$ appropriately, by selecting the best paths of the first step, so as to prolong the network lifetime.

\subsection{Computing sets of paths with acceptable delay} \label{sec::ComputePathSets}

The first step of the method is algorithm \ref{algo::ComputePathSets}, which we call \texttt{ComputePathSets} algorithm (or \texttt{CPS} in short). The output of this algorithm is two sets of paths for each data piece $d \in D$. Specifically, the first path set, denoted as $\mathbf{\Omega}_{pc_d}$, includes the $k$ shortest data propagation paths (where $k$ is a predefined constant - this is a choice that reduces the number of hops, thus limiting the total number of flows insisting on any given link) between each cache node $p$ and each consumer $c_d$, while at the same time respecting the data access delay constraint. The second path set, denoted as $\mathbf{\Omega}_{ps_d}$, includes the $k$-shortest paths between each cache node $p$ and each producer $s_d$.

The \texttt{CPS} algorithm works as follows. In the beginning (line \ref{algo::init}), the algorithm runs  an initialization phase, in which it assigns a value to the $l^{(h)}$ variable, which represents an accurate estimation of the one-hop delay values in the network. In order to derive an appropriate value for $l^{(h)}$, the network controller measures different one-hop data propagation delays within the industrial edge deployment, and gathers a sufficiently representative dataset of $l_{uv}$ delay measurements from different pairs of nodes $u,v \in V$. The algorithm selects the highest measured value for assignment to $l^{(h)}$ (or, in other words, the worst-case one-hop delay). The worst-case delay on each link could alternatively be stored, and use those values to compute the expected delay on the paths, improving the accuracy of the algorithm, without changing the computational complexity of the algorithm. Then (lines \ref{algo::ford}-\ref{algo::forp}), for each data piece $d \in D$, the algorithm examines each cache node $p \in P$ and finds the set $\mathbf{\Omega}_{pc_d}$ of the $k$-shortest paths between $p$ and the consumer $c_d$ (line \ref{algo::pathsc}), and the set $\mathbf{\Omega}_{ps_d}$ of the $k$-shortest paths between $p$ and the source $s_d$ (line \ref{algo::pathss}), in terms of data access delays $L_{p_dc_d}$ and $L_{s_dp_d}$. Note that the $k$-shortest paths can be found using well-known algorithms \cite{eppstein}, \cite{yen}, and that in the case of the path between $p$ and $c_d$ (line \ref{algo::pathsc}), the algorithm excludes the paths resulting in a delay more than $L_{\text{max}}$, thus guaranteeing the constraint imposed by equation \ref{eq::constr}. The time complexity of \texttt{CPS} algorithm is dependent on the algorithm used for computing the $k$-shortest paths (lines \ref{algo::pathsc}-\ref{algo::pathss}). In the case where we use Yen's algorithm \cite{yen}, we have a pseudo-polynomial time complexity of $\mathcal{O}(dp(kn(m + n\log n)))$, where $n = |V|, m = |E|, p = |P|, d = |D|$.

\begin{algorithm}[t!]
\DontPrintSemicolon
\SetKwInOut{Input}{Input}\SetKwInOut{Output}{Output}
 	$l^{(h)}$: assign value through initialization phase \label{algo::init} \;
 \Input{$G, P, D, l^{(h)}, L_{\text{max}}, k$}
	\ForEach{$d \in D$}{ \label{algo::ford}
		\ForEach{$p \in P$}{ \label{algo::forp}
			$\mathbf{\Omega}_{pc_d} \overset{\text{\cite{eppstein}, \cite{yen}}}{=} \left\{ \Omega^{(1)}_{pc_d}, ...,  \Omega^{(k)}_{pc_d}\right\} \setminus \left\{ \Omega^{(j)}_{pc_d} \mid  l^{(h)} \cdot |\Omega^{(j)}_{pc_d}|  > L_{\text{max}}, \forall j: 1 \leq j \leq k \right\}$, $\forall c_d \in V$ \label{algo::pathsc} \;
			$\mathbf{\Omega}_{ps_d} \overset{\text{\cite{eppstein}, \cite{yen}}}{=} \left\{ \Omega^{(1)}_{ps_d}, ...,  \Omega^{(k)}_{ps_d}\right\}$, $\forall s_d \in V$ \label{algo::pathss} \;
			
		}
	}
\Output{$\mathbf{\Omega}_{pc_d}, \mathbf{\Omega}_{ps_d}$}
 \caption{\texttt{ComputePathSets}}
 \label{algo::ComputePathSets}
\end{algorithm}

\subsection{Enabling data caching for lifetime prolongation} \label{sec::DataCacheAccess}

After running \texttt{CPS} algorithm, the network controller is aware of the delay bounded data distribution options for each data piece. The second step of the method is algorithm \ref{algo::DataCacheAccess}, which we call \texttt{DataCacheAccess} (or \texttt{DCA} in short). The rationale behind \texttt{DCA} is to prioritize the distribution of the more frequently requested data pieces, due to the fact that they will require more data propagations in the network and consequently will use more energy. The output of this algorithm is the exact configuration of variables $x^{c_d}_{uv}, x^{s_d}_{uv}$, which in turn regulates the data distribution and caching process in the network.

The \texttt{DCA} algorithm works as follows. In the beginning (line \ref{algo2::sort}), it sorts the set $D$ which contains the data pieces, according to their data consumption rate, resulting in the totally ordered set $(D, \leq)$. Then (line \ref{algo2::ford}), starting from the data piece which necessitates the highest paths activation rate, and for each data piece $d$, it computes which are the paths that guarantee the longest lifetime, both for the caching process between $s_d$ and a potential $p$, and for the data access between the same $p$ and $c_d$ (lines \ref{algo2::runproxy}-\ref{algo2::finishproxy}). This is achieved by examining for each data cache $p$ (line \ref{algo2::runproxy}) all the available path combinations between the different paths provided in the two output sets of \texttt{CPS} algorithm, $\mathbf{\Omega}_{pc_d}, \mathbf{\Omega}_{ps_d}$ (line \ref{algo2::checksets}), and by checking for each path combination, the resulting lifetime $T_u(\mathbf{x})$, for each $u$ included in the paths. Finally, \texttt{DCA} configures appropriately the variables $x^{s_d}_{uv}$ and $x^{c_d}_{uv}$, and updates the new lifetime values for every $u$ included in the selected paths, so that the paths for the next data piece $d$ can be computed given the new network lifetime value. The time complexity of \texttt{DCA} is $\mathcal{O}(dpk^2)$ for each path, where $p = |P|, d = |D|$.

\begin{algorithm}[t!]
\DontPrintSemicolon
\SetKwInOut{Input}{Input}\SetKwInOut{Output}{Output}
 \Input{$P, D, E_u, \epsilon_{uv}, \mathbf{\Omega}_{pc_d}, \mathbf{\Omega}_{ps_d}$}
	$(D, \leq) :$ ordered $D$ from highest to lowest \label{algo::sortdata} $r^c_d$ \label{algo2::sort} \; 
	\For {$d = (D, \leq)[1]:1:(D, \leq)[|D|]$}{ \label{algo2::ford}
			$\textnormal{maxlife} = 0$\;
		\ForEach {$p \in P \textnormal{ with } \mathbf{\Omega}_{pc_d} \neq \varnothing$ and $\mathbf{\Omega}_{ps_d} \neq \varnothing$}{ \label{algo2::runproxy}
			 \For {$i=1:1:k$}{
			 	\For{$j=1:1:k$}{
			 		\If{$T_u(\mathbf{x}) > \textnormal{maxlife}, \forall u \in \Omega^{(i)}_{ps_d} \cup \Omega_{pc_d}^{(j)}$}{ \label{algo2::checksets}
						$\Pi_{s_d} = \Omega^{(i)}_{ps_d}$ \;
						$\Pi_{c_d} = \Omega_{pc_d}^{(j)}$ \;
						$\textnormal{maxlife} = T(\mathbf{x})$
					}
				}
 			}
		}\label{algo2::finishproxy}
		$x^{s_d}_{uv} = 1, \forall (u,v) \in \Pi_{s_d}$\;
		$\texttt{update}(T_u(\mathbf{x})), \forall u \in \Pi_{s_d}$\;
		$x^{c_d}_{uv} = 1, \forall (u,v) \in \Pi_{c_d}$\;
		$\texttt{update}(T_u(\mathbf{x})), \forall u \in \Pi_{c_d}$\;
	}
\Output{$x^{c_d}_{uv}, x^{s_d}_{uv}$}
\texttt{activate}$(x^{c_d}_{uv}, x^{s_d}_{uv})$
 \caption{\texttt{DataCacheAccess}}
 \label{algo::DataCacheAccess}
\end{algorithm}

\subsection{An optimal benchmark} \label{sec::opt}

In order to demonstrate the performance of our algorithm, we would ideally prefer a benchmark optimization technique which could optimally solve the problem. However, due to the fact that \texttt{DDA} is an $\mathcal{N}\mathcal{P}$-complete problem (Theorem \ref{theo::theo}), we are unable to provide such a solution. In order to overcome this issue, we provide a technique that is able to solve a relaxed formulation of the \texttt{DDA} problem. We use the performance of this technique as a benchmark, in order to have a performance upper bound for \texttt{DDA}. The technique employed is the formulation of the relaxed version of the problem as an integer program without considering the delay constraint, and a further relaxation which enables the ability to distribute a single data piece through multiple edges, rendering the solution into a solvable linear program. The integer program can be formulated as follows. Note that the integer constraints enforce the variables in the optimization problem to take on only integer values.

The objective function (\ref{problem::of}) remains the same with the one defined in equation \ref{eq::obj}, which is the maximization of the time until the first node in the network depletes its energy. Constraints (\ref{problem::flowstd}-\ref{problem::flowin}) guarantee the data flow conservation for all nodes, including the data producers and consumers. Constraints (\ref{problem::energy}) guarantee that the energy consumption of each node $u$ will not exceed the initially available energy value $E_u$. Constraints (\ref{problem::limit}) guarantee that each data piece is propagated from $u$ through one and only one edge $(u,v)$. However, although the variables $x^{s_i}_{uv}$ and $x^{c_i}_{uv}$ are initially set to be integers, according to the formulation of section \ref{sec::obj}, in order to be able to provide a feasible solution for this integer program, we consider a further simplification, the linear relaxation of the program, by enforcing the weaker constraints $x^{s_i}_{uv}, x^{c_i}_{uv}  \in [0,1]$ (\ref{problem::xbinary}). This relaxation leads to a linear multi-commodity flow problem with fractional flows, which in our case means that an arbitrary fraction of the data piece is sent through edge $(u,v)$ and thus a data piece can be broken in smaller pieces of any fractional size, and sent across multiple links. This fact does not allow us to evaluate its performance in real settings, but only in simulations, however, in all cases, the performance of the linear program solution is at least as good as the performance of a solution for the integer program, because any integer program solution would also be a valid linear program solution. Note that if we want to ensure that at least one cache node is included at the distribution of the data pieces, we can impose an additional set of constraints which ensure that $\sum_{p \in P} x^{c_i}_{pv} \geq 1, \forall i$. 

\begin{align}
\text{maximize: } T(\mathbf{x}) \label{problem::of}
\end{align}
subject to:
\begin{align}
&  \sum_{v \in V} (x^{s_i}_{uv} + x^{c_i}_{uv} - x^{s_i}_{vu} - x^{c_i}_{vu}) = 0 & \forall u \in V \setminus \{s_i, c_i\} \label{problem::flowstd}\\
& \sum_{v \in V} (x^{s_i}_{s_iv} - x^{s_i}_{vs_i}) = 1 & \forall s_i \in V \label{problem::flowout}\\
& \sum_{v \in V} (x^{c_i}_{s_iv} - x^{c_i}_{vs_i}) = 0 & \forall s_i \in V \label{problem::flowout1}\\
& \sum_{v \in V} (x^{c_i}_{c_iv} - x^{c_i}_{vc_i}) = -1 & \forall c_i \in V \label{problem::flowin1}\\
& \sum_{v \in V} (x^{s_i}_{c_iv} - x^{s_i}_{vc_i}) = 0 & \forall c_i \in V \label{problem::flowin}\\
& \sum_{v \in V} \sum_{i} \epsilon_{uv}  (r^g_i x^{s_i}_{uv} + r^c_i x^{c_i}_{uv}) \leq E_u & \forall u \in V \label{problem::energy}\\
& \sum_{v \in V} x^{s_i}_{uv} \leq 1, \sum_{v \in V} x^{c_i}_{uv} \leq 1 & \forall u \in V, \forall i \label{problem::limit}\\
&  x^{s_i}_{uv}, x^{c_i}_{uv} \in [0,1] & \forall u, v \in V, \forall i \label{problem::xbinary}
\end{align}

\subsection{Experimental evaluation of the algorithm} \label{sec::exps}

In order to demonstrate the performance of the algorithm in an environment with real devices, as well as to validate the simulation model that we developed for the larger scale experiments (presented later, at section \ref{sec::sims}), we conduct an experimental implementation and evaluation. For the purposes of the experiments, we used the Euratech testbed of FIT IoT-LAB platform situated in Lille, France. IoT-LAB is a large-scale collection of open wireless IoT testbeds, operated by the French CNRS and INRIA research institutions \cite{7389098}. For acquiring the results of the linear program and the performance upper bound, we used the Matlab \texttt{linprog} solver. Among all supported operating systems (e.g., TinyOS, OpenWSN, Contiki), we conducted our implementation on TinyOS. The details of the experimental setup are exposed in table \ref{tab::parameters}.

\begin{figure}[t!]
\centering
    \begin{subfigure}[b]{0.24\textwidth}
    \centering
        \includegraphics[width=\columnwidth]{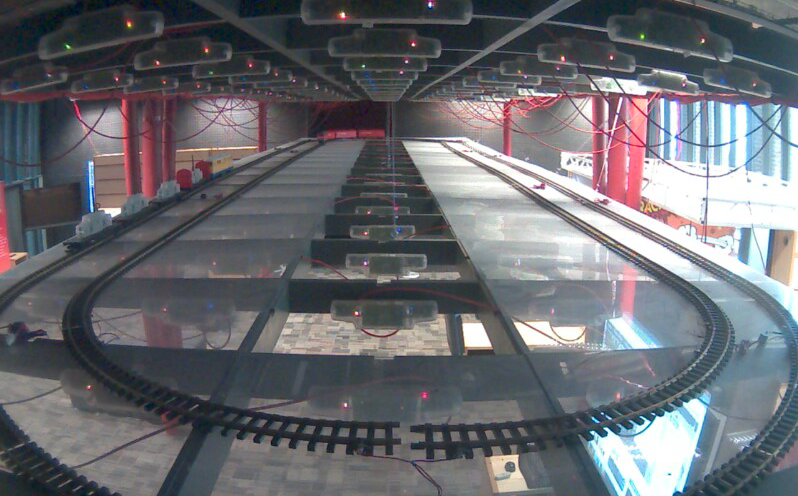}
        \caption{Euratech testbed.}
        \label{fig::testbed}
    \end{subfigure}
    \begin{subfigure}[b]{0.24\textwidth}
    \centering
        \includegraphics[width=\columnwidth]{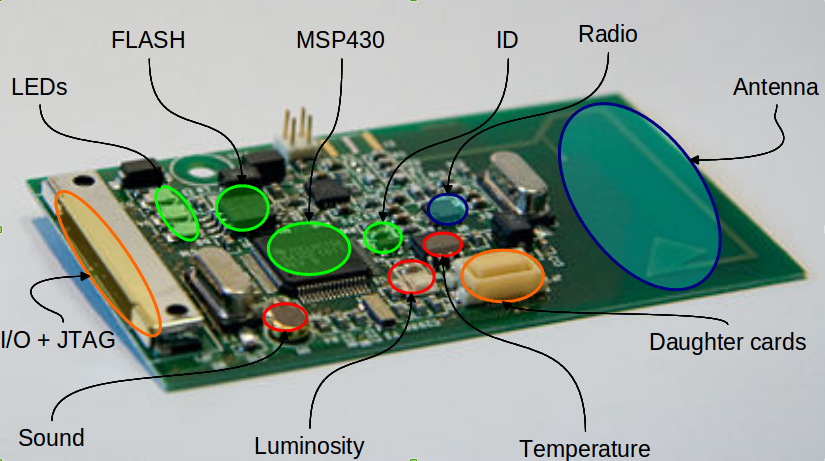}
        \caption{WSN430 open node.}
        \label{fig::wsn430}
    \end{subfigure}    
          
    \begin{subfigure}[b]{0.24\textwidth}
    \centering
        \includegraphics[width=\columnwidth]{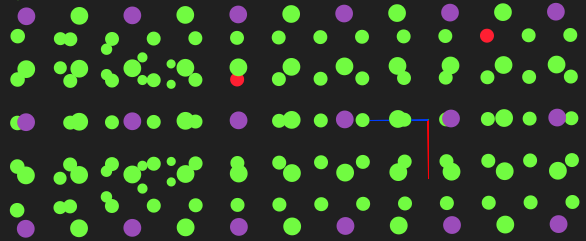}
        \caption{Nodes reservation.}
        \label{fig::reservation}
    \end{subfigure}
    \begin{subfigure}[b]{0.24\textwidth}
        \centering
        \includegraphics[width=\columnwidth,trim={0 2.5cm 0 2cm},clip]{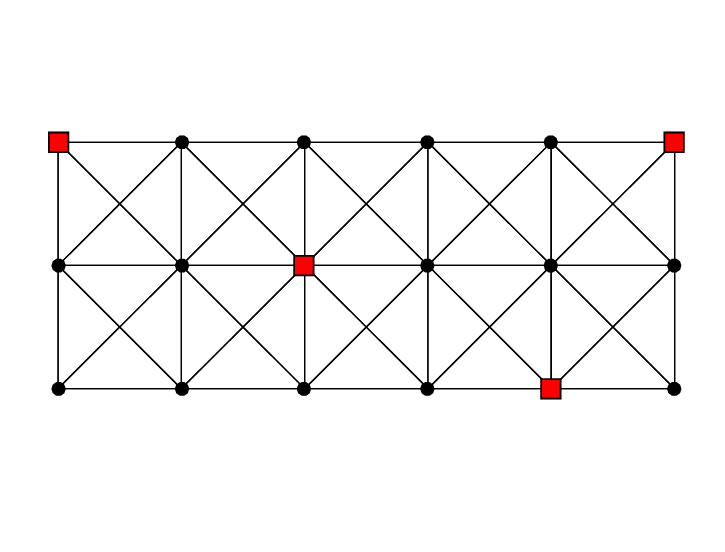}
        \caption{Network topology.}
        \label{fig::network-real}
    \end{subfigure}
    \caption{Experimental setup.}\label{fig::experiments}
\end{figure}

\emph{Topology.} The Euratech testbed was deployed in the Inria Lille - Nord Europe showroom and $224$ nodes were deployed as follows: two horizontal layers in grid formation of $5\times19$ nodes each and $34$ nodes attached to a wall, at a distance of $0.60$ m to each other. Fig.~\ref{fig::testbed} displays a photo of the testbed. Due to the fact that this is a dense deployment, and in order to acquire a realistic indoor industrial topology we reserved $18$ of the testbed's nodes (purple dots in Fig.~\ref{fig::reservation}); a selection which results in a sparser topology with node distances of $1.2 - 1.7$ m. Moreover, we observe that even if this topology is sparser, the resulting multi-hop network can still be considered dense, depending on the output power level of the nodes. Therefore, after having investigated various output power levels (Fig.~\ref{fig::levels}), we targeted a transmission power of $3$ m, with $\gamma = 0.6$, which in turn results to a neighborhood with $5$ neighboring nodes on average, where $v \in N_u$ when $\delta(u,v) \leq 2$. We selected $4$ nodes at random to act as cache nodes in the network (so as to maintain the modeling assumption $|P| \ll |V-P|$). Given this configuration, we obtain a topology which is depicted in Fig.~\ref{fig::network-real}.

\begin{table}[t!]
\centering
\caption{Experimental parameters.}
\begin{tabular}{ l | l }\hline
\textbf{Parameter} & \textbf{Value}\\\hline
\multicolumn{2}{c}{\textbf{Topology}} \\\hline
deployment dimensions (2D grid) & $2.4$ m $\times$ $6.0$ m\\
number of nodes $|V|$, edges $|E|$, caches $|P|$ & $18, 47, 4$\\
node distances $\delta (u,v)$ & $1.2 - 1.7$ m\\
transmission range $\rho_u$, parameter $\gamma$ & $3$ m, $0.6$\\
neighborhood $N_u$ & $v$, with $\delta(u,v) \leq 2$ m\\\hline
\multicolumn{2}{c}{\textbf{Hardware}} \\\hline
MCU (ultra low-power) & MSP430   \\
antenna (IEEE 802.15.4)  & CC2420  \\
max. battery capacity & $830$ mAh, $3.7$ V\\
node energies $E_u$, cache energies $E_p$ & $0-1, 3$ Wh\\
transmission power & $-25$ dBm\\\hline
\multicolumn{2}{c}{\textbf{Time}} \\\hline
time cycle $\tau$ & $1$ sec \\
delay threshold $L_{\text{max}}$ & $120$ ms \\
max. measured one-hop delay $l^{(h)}$& $28$ ms\\
experiment duration & $20$ min\\\hline
\multicolumn{2}{c}{\textbf{Data}}\\\hline
percentage of consumers $c_d$ & $0.05 - 45\%$\\
generation/consumption rates $r^g_d,r^c_d$ & $1-8$ $d$/sec\\
data piece size (incl. headers and CRC) & $9$ bytes\\\hline
\end{tabular}
\label{tab::parameters}
\end{table}

\emph{Hardware.} For the experiments we use the WSN430 open nodes, the design of which is displayed in Fig.~\ref{fig::wsn430}. The WSN430 open node is a mote based on a low power MSP430-based platform, with a set of standard sensors and IEEE 802.15.4 radio interface at 2.4 GHz, using a CC2420 antenna \cite{cc2420}. We configured the antenna TX power at $-25$ dBm and, according to the CC2420 antenna datasheet \cite{cc2420}, we acquire the preferred range $\rho_u = 3$ m. The nodes are battery operated with maximum capacity of $830$ mAh at $3.7$ V. We equip the nodes with $0-1$ Wh and the cache nodes with $3$ Wh of energy.

\emph{Time.} In order to perform the experiments in the most realistic way, we align the $L_{\text{max}}$ value with the official requirements of future network-based communications for Industry 4.0, for the targeted industrial applications. Both the \emph{WG1 of Plattform Industrie 4.0} (reference architectures, standards and norms) \cite{reqs} and the \emph{Expert Committee 7.2 of ITG} (radio systems) \cite{itg} set the delay requirements for condition monitoring applications to around $100$ ms, so we set the data access delay threshold to $L_{\text{max}} = 120$ ms. We set the time cycle $\tau = 1$ sec and we assigned $l^{(h)} = 28$ ms after the initialization phase, as explained in the next subsection. Although the experiments duration was set to $20$ minutes (which is a very short period to demonstrate the industrial operation effects on the network lifetime) due to resource sharing restrictions with other users of the FIT IoT-LAB platform, we extended our results to longer periods of time, based on the measurements we obtained.

\emph{Data.} We variate the percentage of data sources and consumers between the values of $0.05 - 45\%$ on the network nodes number, with diverse data piece generation and consumption rates of $r^g_d$ and $r^c_d$ between $1$ and $8$ $d$/sec. The data piece size was set to $9$ bytes, which corresponds to all necessary information for routing and application operations (nodeID, sensed values, etc.).

\subsection{Experimental results}

\begin{figure}[t!]
\centering
    \begin{subfigure}[b]{0.24\textwidth}
    \centering
        \includegraphics[width=\columnwidth]{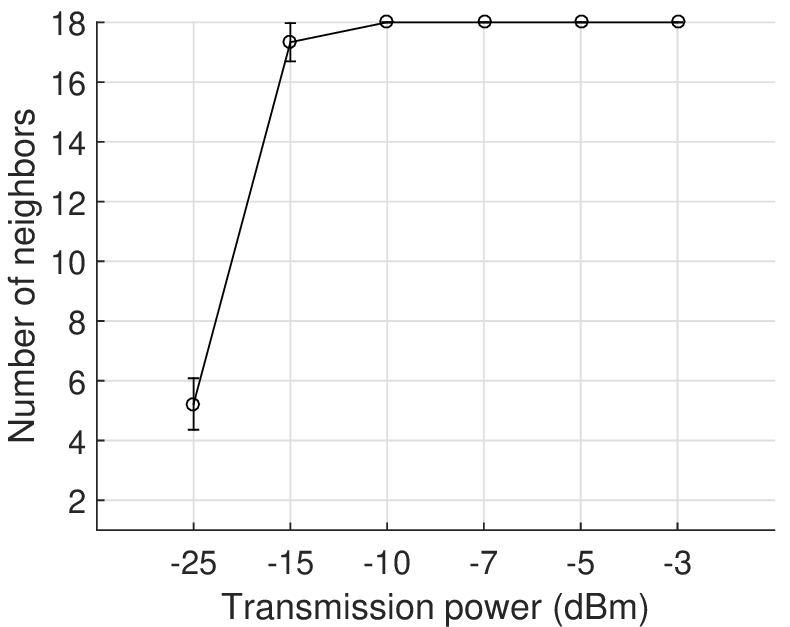}
        \caption{Average size of $N_u$ for different TX power levels and $\gamma = 0.6$.}
        \label{fig::levels}
    \end{subfigure}    
    \begin{subfigure}[b]{0.24\textwidth}
    \centering
        \includegraphics[width=\columnwidth]{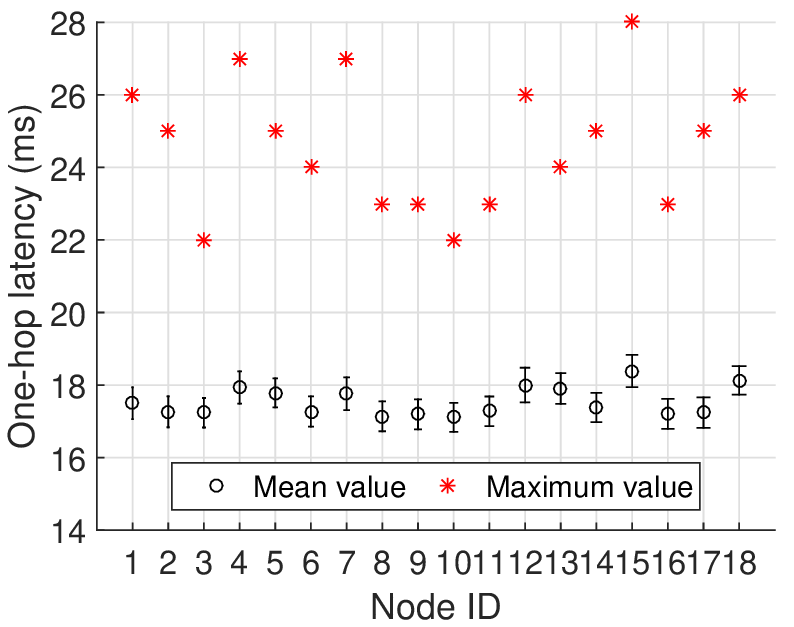}
    \caption{Mean and maximum one-hop delay measurements.}\label{fig::error}
    \end{subfigure}
    \caption{Initialization measurements.}\label{fig::level-error}
\end{figure}

\emph{Initialization phase}. We ran the initialization phase of \texttt{CPS} (algorithm \ref{algo::ComputePathSets}), so as to assign values to $l^{(h)}$, by measuring times that are needed for propagating a data piece. In order to obtain reliable results, we repeated the propagation measurements multiple times for all $18$ nodes, for different pairs of transmitting and receiving nodes of Euratech testbed. We concluded to the measurements that are shown in the Fig.~\ref{fig::error} (mean and maximum values), after measuring the relevant delays using WSN430 with CC2420 and TinyOS. We can see that the delay values of data propagation from one node to another significantly vary. While the mean delays are $17-19$ ms, the maximum propagation delays observed are $22-28$ ms. In order to provide maximum delay guarantees with respect to $L_{\text{max}}$, we assign to $l^{(h)}$ the maximum value observed in the experiments, $l^{(h)} = 28$ ms.

\begin{figure}[t!]
\centering
    \begin{subfigure}[b]{0.24\textwidth}
    \centering
        \includegraphics[width=\columnwidth]{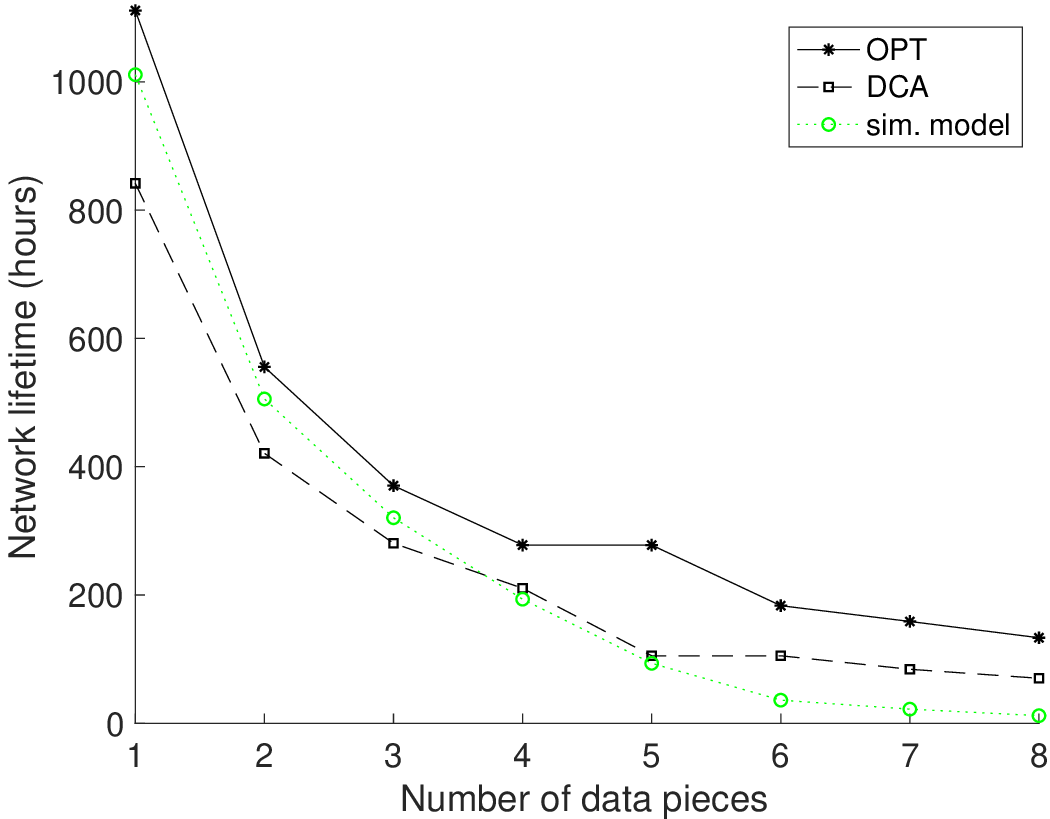}
        \caption{Network lifetime.}
        \label{fig::lifetime}
    \end{subfigure}
    \begin{subfigure}[b]{0.24\textwidth}
    \centering
        \includegraphics[width=\columnwidth]{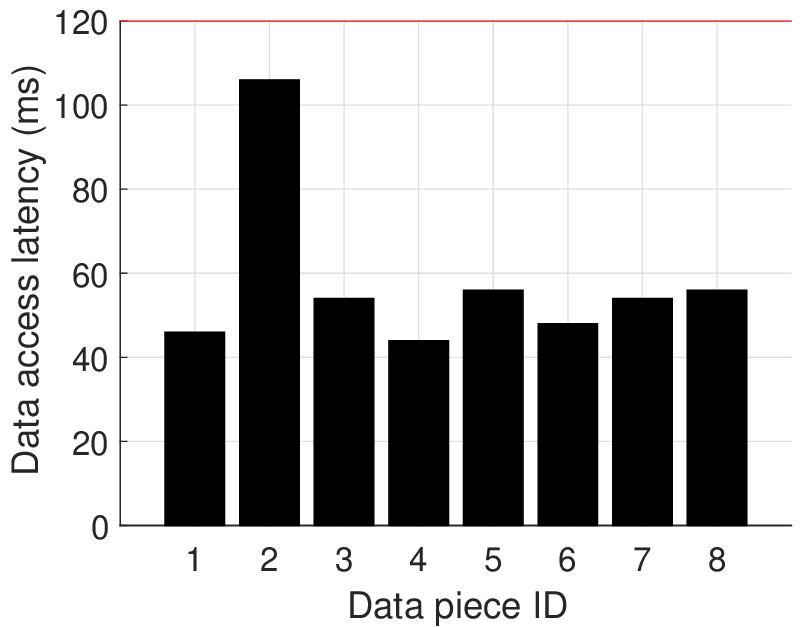}
        \caption{\texttt{DCA} data access delay.}
        \label{fig::latency}
    \end{subfigure}
    \begin{subfigure}[b]{0.3\textwidth}
        \centering
        \includegraphics[width=0.6\columnwidth]{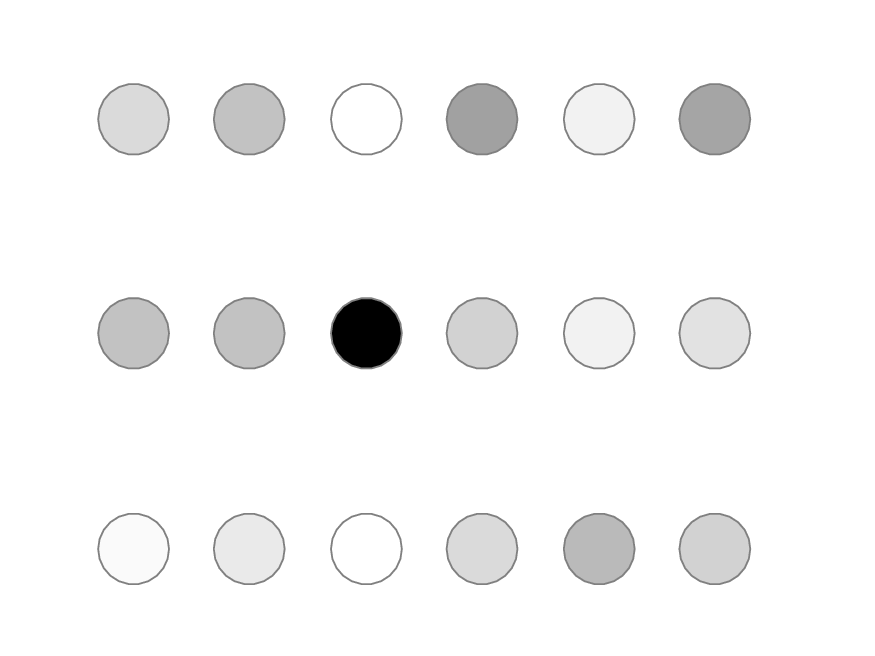}
        \caption{\texttt{PDD} energy balance, $d = 8$.}
        \label{fig::energymap-ours}
    \end{subfigure}
    \caption{Experimental results.}\label{fig::results}
\end{figure}

\emph{Network lifetime.} In order to experimentally demonstrate the quantitative approximation of the performance of \texttt{DCA} (algorithm \ref{algo::DataCacheAccess}) to the optimal performance, we ran the \texttt{DCA} defined data distribution in the network for increasing number of consumers ($c_d = 1-8$) and for increasing data piece generation and consumption rates ($r^g_d, r^c_d = 1-8$ $d$/sec), and we compared the results to the performance of the theoretically optimal solution, i.e., the solution of the relaxed problem shown in section \ref{sec::opt}. The comparison is shown in Fig.~\ref{fig::lifetime}. We can see that \texttt{DCA} approximates the performance of the optimal solution with a difference of $70-300$ hours in terms of network lifetime. Considering the many relaxations that the optimal solution assumes (see Section \ref{sec::opt}), this is a remarkably good result (the optimal solution is computed over much milder constraints, and therefore it is more a computable benchmark than an optimal solution). 

\emph{Data access delay.} Fig.~\ref{fig::latency} depicts the data access delay achieved by each one of the consumers in the network. We measured the delays asynchronously by individual requests of the consumers to the corresponding cache nodes which cached their data. We can see that the delay values lie below the data access delay threshold (red line in Fig.~\ref{fig::latency}). Note that the consumer requesting data piece $2$ has a longer access time because its position in the network is farther from a cache node, compared to the positions of the rest of the nodes.

\emph{Energy balance.} For the \texttt{DCA} case, we measured the energy dissipated on each individual node, in order to get an indication about the energy consumption balance over the network. More specifically, we present graphically the spatial evolution of energy consumption in the network. Nodes with high energy consumption are depicted with dark colors. In contrast, nodes with low energy consumption are depicted with bright colors. As exposed in Fig.~\ref{fig::energymap-ours}, \texttt{DCA} achieves a balanced energy consumption over the network, having chosen to move data through the nodes with high initial energy supplies. The reason why there is a significantly darker node in the center is because, as shown in Fig.~\ref{fig::network-real}, it is a central cache node node with $8$ edges which serves a lot of requests, and lies on numerous paths when $|D|=8$.

\emph{Validation of the simulation model.} In order to obtain some conclusions regarding the performance in a larger scale of the methods presented in this paper, we developed a simulation environment in Matlab, modeled and based on the real world parameters of the wireless edge network. The simulation environment is an implementation of the model presented in section \ref{sec::basics}, and was tested and validated against the experimental results, under the same network parameters (table \ref{tab::parameters}), as shown in Fig.~\ref{fig::lifetime} (green line). We can see that the results obtained from the simulations approximate in a satisfying way the results obtained by the experiments. The simulation curve is naturally smoother than the experimental curve, due to the fact that in real world experiments various phenomena can variate the behavioral characteristics of wireless communication. The results of Fig.~\ref{fig::lifetime} also validate the fact that, for the considered scenario where the network is carefully deployed, the graph modeling assumptions of section \ref{sec::basics} are a fairly good choice (the abstraction of a wireless network as a graph is debatable, mainly as the links between the nodes can be quite dynamic. This has been largely debated in community and as a result, stochastic geometric models have gained quite a bit of popularity. However, for the considered scenario where the network is carefully deployed this assumption is a fairly good one).

\section{Dynamic Path Reconfiguration Algorithms} \label{sec::impro}

We now go one step beyond the strict, constant data distribution schedule and we provide two online improvements, so that the data distribution can dynamically change before the first node in the network dies: one centralized method which uses local area wireless communication to renew the data distribution schedules according to the current network energy map and one distributed method which periodically rotates the available data distribution paths in a proportionally fair manner. While the first one can optimize paths based on global knowledge, this comes at the cost of additional energy cost for transmission of field nodes status via the local area network. So, it is not immediate to anticipate which solution would be the best one.

\subsection{A centralized improvement using local area wireless}

The first improvement is algorithm \ref{algo::DataCacheAccessPlus} and is called \texttt{DataCacheAccess+} (or \texttt{DCA+} in short). The idea behind this algorithm is that, instead of running the same data distribution schedule (calculated by \texttt{DCA}) until the first node in the network dies, we periodically rerun  \texttt{DCA} with a preselected period of $\alpha\tau$ and apply the new data distribution schedule. This process will keep the network updated at the beginning of each new period with a new, efficient data distribution schedule, based on the latest network energy map. However, this update comes at a cost, since the update for all network nodes is performed via local area wireless communication of the edge nodes to the controller; a way of communication which is much more power hungry than the low-power wireless used in the field (e.g., $15$ dBm for IEEE 802.11 \cite{LM820} instead of $-25$ dBm for IEEE 802.15.4 \cite{cc2420}). 

\texttt{DCA+} works as follows: The algorithm repeats the following process until the first node in the network dies (lines \ref{algo3::repeat}-\ref{algo3::endrepeat}). The network controller runs \texttt{DCA} and calculates the data distribution schedule according to the current energy map of the network (line \ref{algo3::dca}). Then a timer starts counting (line \ref{algo3::timer}), and until the timer reaches the value $\alpha\tau$ (line \ref{algo3::while}), the network is activating the edges provided by the calculated data distribution schedule (line \ref{algo3::act}). After this time point, the edges get deactivated (line \ref{algo3::deact}) and each node in the network communicates its energy status to the network controller via local area wireless (line \ref{algo3::foreach}-\ref{algo3::send}). Then the repeating loop starts over again.

\begin{algorithm}[t!]
\DontPrintSemicolon
\SetKwInOut{Input}{Input}\SetKwInOut{Output}{Output}
 \Input{$P, D, E_u, \epsilon_{uv}, \mathbf{\Omega}_{pc_d}, \mathbf{\Omega}_{ps_d}, \alpha$}
 \Repeat 
 {$\exists u \in V$ with $E_u = 0$}
 { \label{algo3::repeat}
 run \texttt{DataCacheAccess}$(P, D, E_u, \epsilon_{uv}, \mathbf{\Omega}_{pc_d}, \mathbf{\Omega}_{ps_d})$\label{algo3::dca}\;
 start  \texttt{timer}\label{algo3::timer}\;
 \While{\texttt{timer} $< \alpha \tau$ }{\label{algo3::while}
 \texttt{activate}$(x^{c_d}_{uv}, x^{s_d}_{uv})$}\label{algo3::act}

 \texttt{deactivate}$(x^{c_d}_{uv}, x^{s_d}_{uv})$\label{algo3::deact}\;
 \ForEach {$u \in V$}{\label{algo3::foreach}
 send energy status ($E_u$) to network controller via local area wireless communication}\label{algo3::send}
 } \label{algo3::endrepeat}
 \caption{\texttt{DataCacheAccess+}}
 \label{algo::DataCacheAccessPlus}
\end{algorithm}

\subsection{A distributed improvement using online path rotation}

The centralized nature of \texttt{DCA+}, as well as the frequent use of local area wireless communication, leads to an important disadvantage. The data distribution schedule recomputation necessitates additional, significant communication overhead, so that the field devices inform the network controller about their current energy status. This is performed through the local area wireless links, which are significantly more costly than the low power wireless links (Fig.~\ref{fig::arch}), especially in the case of large scale networks, in which many nodes are involved. In order to overcome this issue, we design the distributed algorithm \ref{algo::ProportionallyFairRotation}, which we call \texttt{ProportionallyFairRotation} (\texttt{PFR} in short). \texttt{PFR} acts as a local improvement technique for each data piece, by using all the available paths that \texttt{CPS} has computed, thus solely using low-power wireless communication. The intuition behind \texttt{PFR} is that, for each data piece, all available paths are proportionally rotated according to the initial energy supplies of the node with the least energy in each path. Therefore, \texttt{PFR} is trying to achieve an energetically balanced rotation of all the available paths of each data piece, in order to avoid early node energy depletions.

\begin{algorithm}[t!]
\DontPrintSemicolon
\SetKwInOut{Input}{Input}\SetKwInOut{Output}{Output}
 \Input{$E_u, \mathbf{\Omega}_{pc_d}, \mathbf{\Omega}_{ps_d}$}
	\ForEach{$d \in D$}{
			$\mathbf{\Pi}_d = \begin{pmatrix}
 			\Omega^{(1)}_{1c_d} \cup \Omega^{(1)}_{1s_d} & \Omega^{(2)}_{1c_d} \cup \Omega^{(2)}_{1s_d} & \cdots & \Omega^{(k)}_{1c_d} \cup \Omega^{(k)}_{1s_d} \\
 			\Omega^{(1)}_{2c_d} \cup \Omega^{(1)}_{2s_d} & \Omega^{(2)}_{2c_d} \cup \Omega^{(2)}_{2s_d} & \cdots & \Omega^{(k)}_{2c_d} \cup \Omega^{(k)}_{2s_d} \\
			\vdots  & \vdots & \ddots & \vdots  \\
			\Omega^{(1)}_{pc_d} \cup \Omega^{(1)}_{ps_d} & \Omega^{(2)}_{pc_d} \cup \Omega^{(2)}_{ps_d} & \cdots &  \Omega^{(k)}_{pc_d} \cup \Omega^{(k)}_{ps_d}
			\end{pmatrix}$ \label{line::mat} \;
			$\mathbf{E}_d (i,j) = \min_{u \in \Omega^{(i)}_{jc_d} \cup \Omega^{(i)}_{js_d}} E_u$ \label{line::energ}\;
			$\varepsilon_d = \max_{1 \leq i \leq k, 1 \leq j \leq p} \mathbf{E}_d (i,j)$ \label{line::vare}\;
		
	}
	\Output{$\mathbf{\Pi}_d, \mathbf{E}_d, \varepsilon_d$}
				$i = j = 1$\;
	\Repeat{$\exists u \in V$ with $E_u = 0$}{\label{line::repe}
					start \texttt{timer}$_d$ \label{line::timer}\;
	\ForEach([in parallel]){$d \in D$}{
				\While{\texttt{timer}$_d$ $ < \frac{\mathbf{E}_d (i,j)}{\varepsilon_d} \alpha \tau$}{ \label{line::fair}
					\texttt{activate}$(x^{c_d}_{uv}, x^{s_d}_{uv} \in \mathbf{\Pi}_d (i,j))$ \label{line::activate}
				}
				\texttt{deactivate}$(x^{c_d}_{uv}, x^{s_d}_{uv})$  \label{line::deactivate}\;
				$j++$\;
				\lIf{$j>k$}{
					$i++, j = 1$ \label{line::path1}
				}
				\lIf{$i>p$}{
					$i = 1, j = 1$ \label{line::path2}
				}
			}
}\label{line::repe2}
 \caption{\texttt{ProportionallyFairRotation}}
 \label{algo::ProportionallyFairRotation}
\end{algorithm}

\texttt{PFR} works as follows for each individual data piece $d \in D$: At first it performs a preprocessing step, by creating a matrix including all paths computed by \texttt{CPS} (line \ref{line::mat}), by storing for each path the initial energy level $\mathbf{E}_d (i,j)$ of the node with the smallest initial energy supplies (line \ref{line::energ}), and by holding as reference value the maximum energy level $\varepsilon_d$ of the all minimum energy values (line \ref{line::vare}). The algorithm repeats the following process until the first node in the network dies (lines \ref{line::repe}-\ref{line::repe2}). A local timer starts counting (line \ref{line::timer}) in parallel for each data piece in a proportionally fair way, in the sense that the timer duration lasts for an amount of time proportional to the path's least energy ratio to the maximum value $\varepsilon_d$ of the minimum values of all paths assigned to $d$ (line \ref{line::fair}). The value of $\varepsilon_d$ will serve as the reference value so as to derive the duration of the rotation timer. The timer formulation imposes a periodicity on each path of a fraction of $\alpha \tau$ ($0 <  \frac{\mathbf{E}_d (i,j)}{\varepsilon_d}  \leq 1$).  As long as the timer is active, the network is activating the edges of the current data distribution path (line \ref{line::activate}). After this time point, the edges get deactivated (line \ref{line::deactivate}) and the data piece starts getting distributed by the next path. Then the repeating loop starts over again. The next path is selected serially from the matrix $\mathbf{\Pi}_d$, by exhausting all the $k$ paths the given cache node and then moving to the next cache node (lines \ref{line::path1}-\ref{line::path2}). Although this is a serial process based on the position of the paths in matrix $\mathbf{\Pi}_d$, there is no provisioning for the positioning and the selection of the first path to be chosen and thus the overall path selection is a random process. There is no guarantee for maintaining nodes in the network alive. Consequently, a node might run out of energy at any given time, and at this point we measure the lifetime (line \ref{line::repe2}). While, in general, we expect that distributed solutions like \texttt{PFR} can be more energy efficient than centralized solutions (distributed approaches use energetically cheaper wireless technologies), in our case, the extent of those energy gains is not a straightforward guess. For this reason, he highlight and demonstrate the performance of \texttt{PFR} against the performance of \texttt{DCA} and \texttt{DCA+}.

\begin{figure*}[t!]
\centering
    \begin{subfigure}[b]{0.24\textwidth}
    \centering
        \includegraphics[width=\columnwidth]{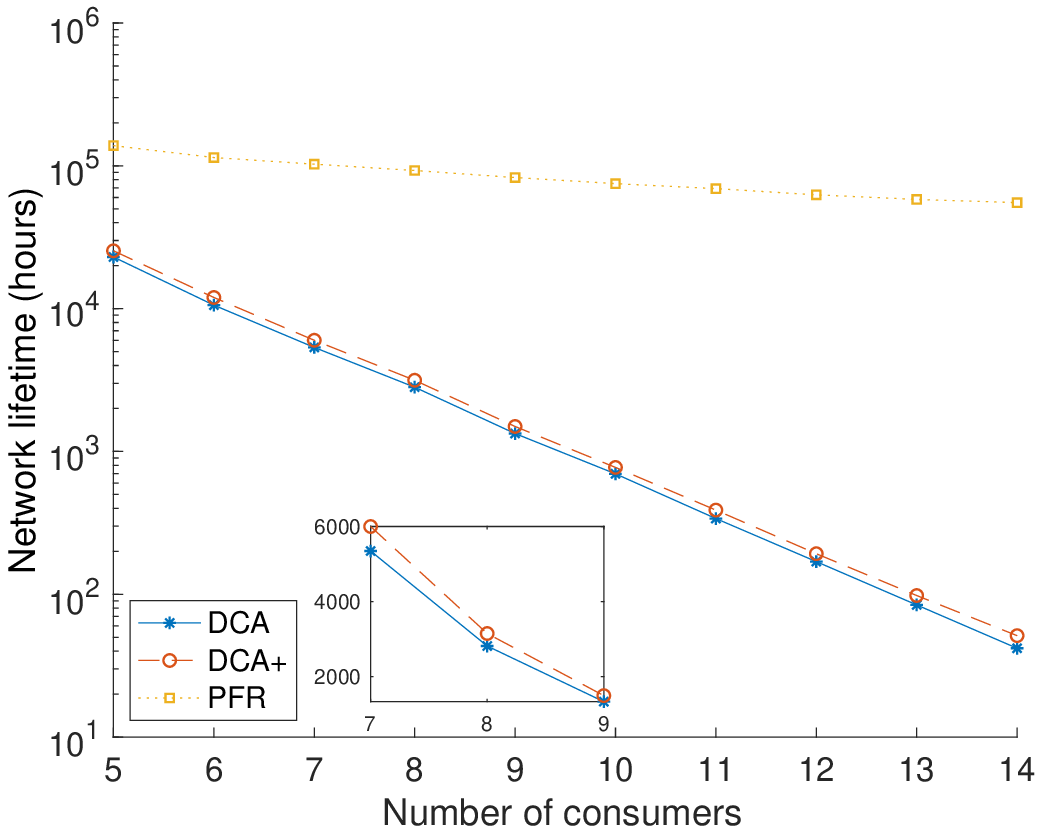}
        \caption{25 nodes.}
        \label{fig::dca1}
    \end{subfigure}
    \label{fig::dca-comp}
    \begin{subfigure}[b]{0.24\textwidth}
    \centering
        \includegraphics[width=\columnwidth]{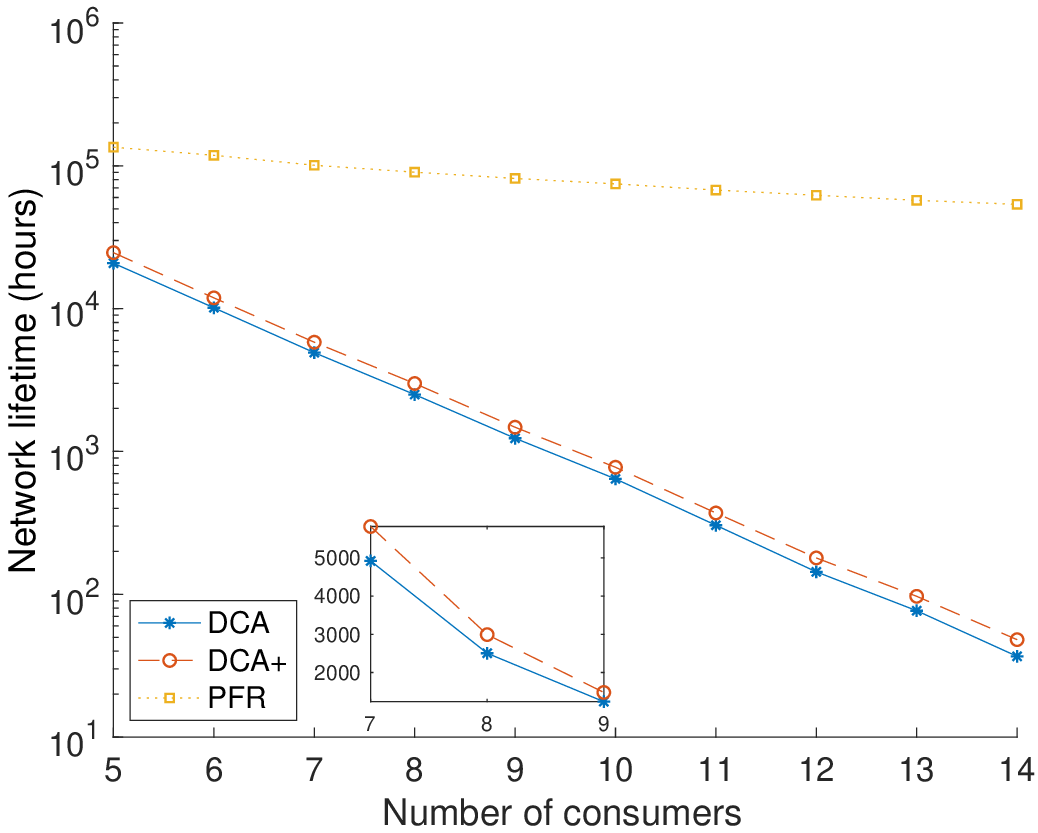}
        \caption{36 nodes.}
        \label{fig::dca2}
    \end{subfigure}
    \begin{subfigure}[b]{0.24\textwidth}
    \centering
        \includegraphics[width=\columnwidth]{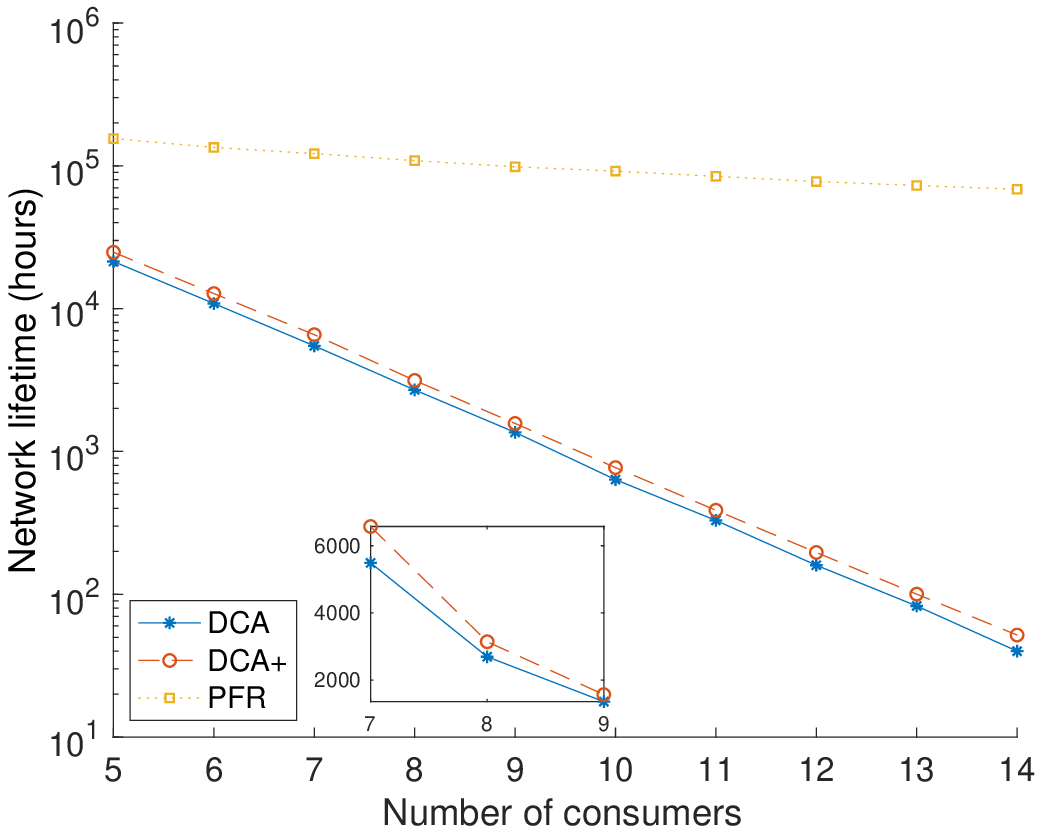}
        \caption{49 nodes.}
        \label{fig::dca3}
    \end{subfigure}
    \begin{subfigure}[b]{0.24\textwidth}
    \centering
        \includegraphics[width=\columnwidth]{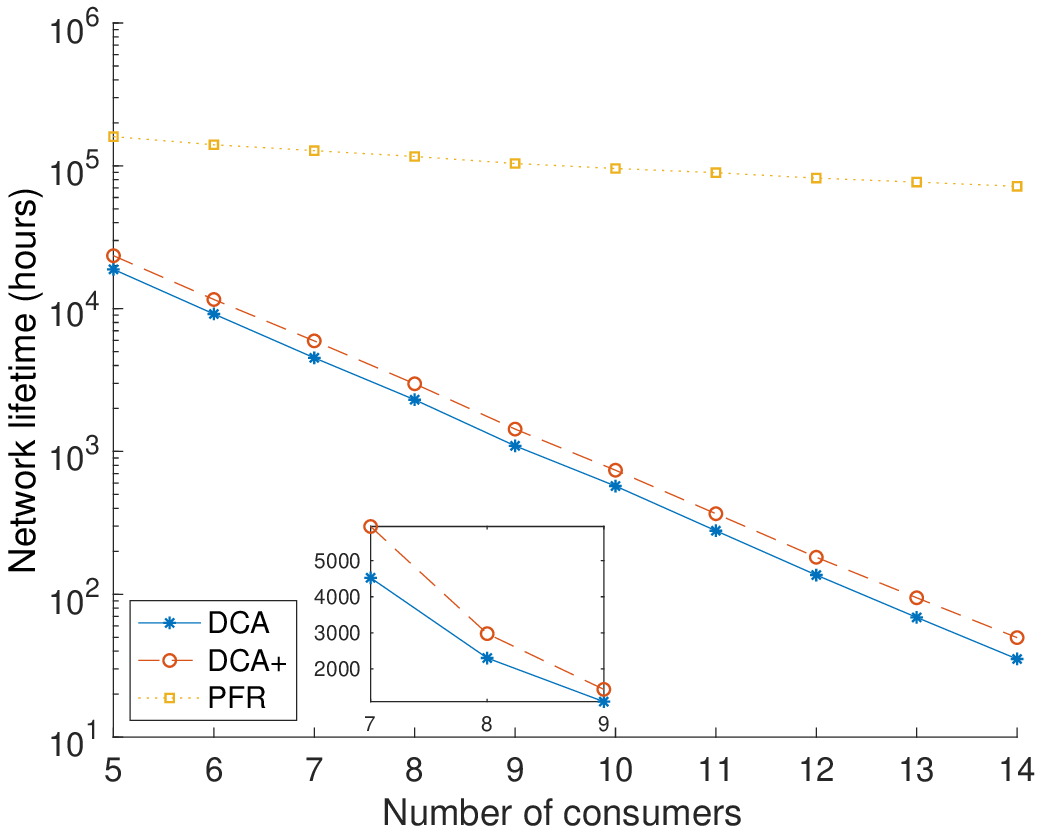}
        \caption{64 nodes.}
        \label{fig::dca4}
    \end{subfigure}
    
   \begin{subfigure}[b]{0.24\textwidth}
    \centering
        \includegraphics[width=\columnwidth]{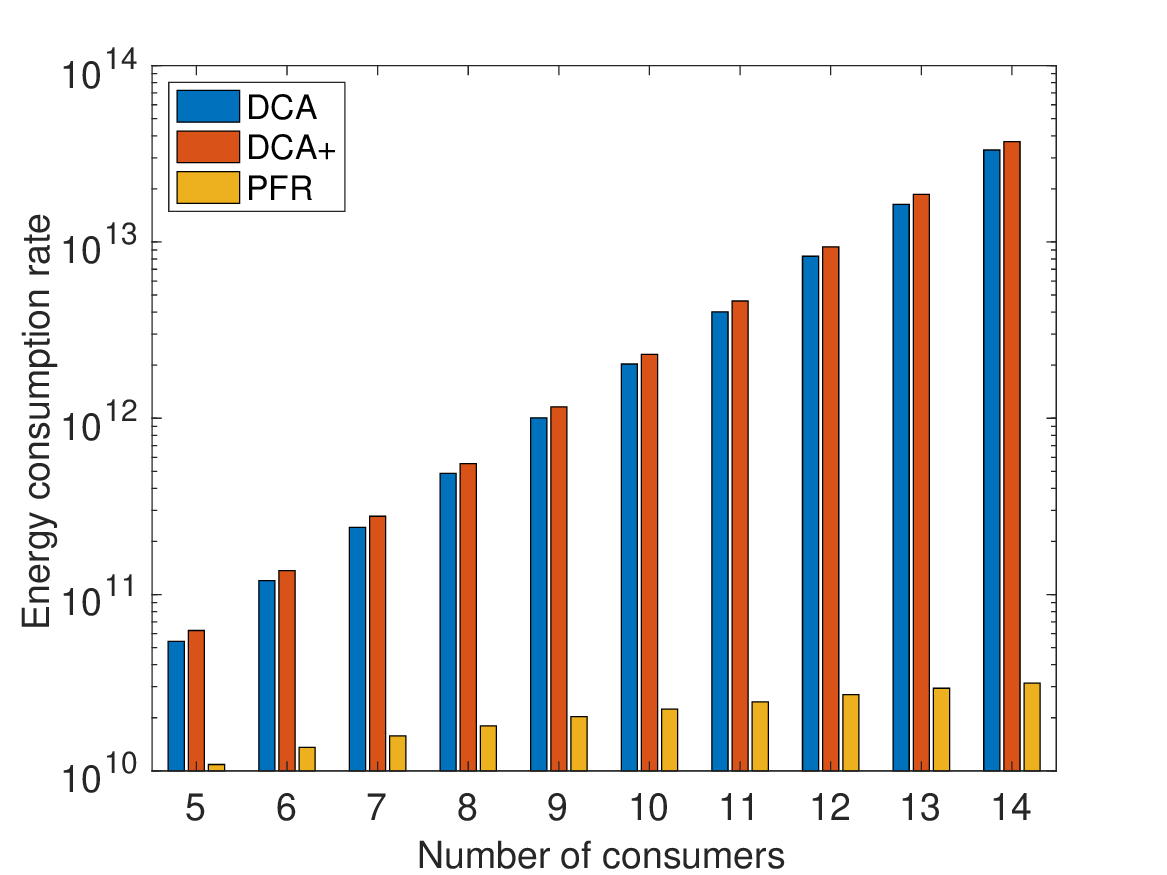}
        \caption{25 nodes.}
        \label{fig::dca1-nrg}
    \end{subfigure}
    \label{fig::dca-comp}
    \begin{subfigure}[b]{0.24\textwidth}
    \centering
        \includegraphics[width=\columnwidth]{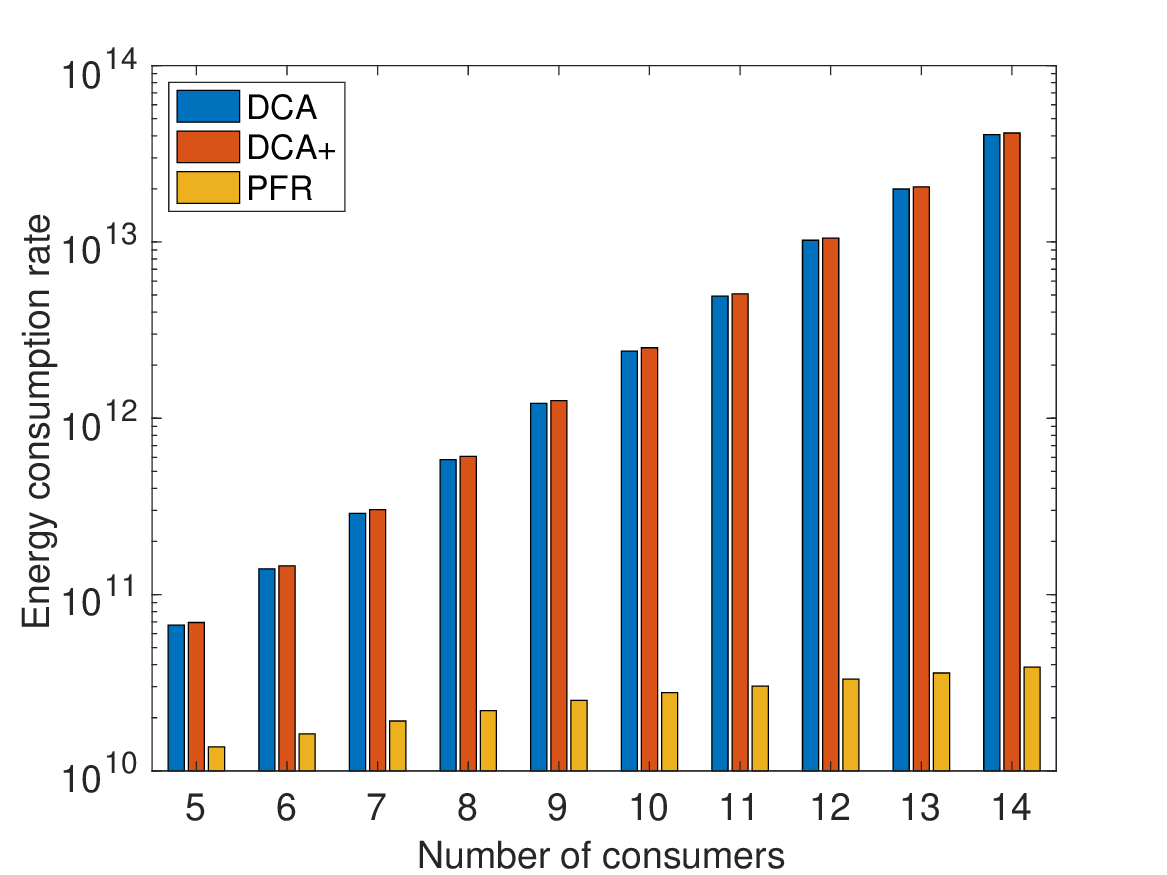}
        \caption{36 nodes.}
        \label{fig::dca2-nrg}
    \end{subfigure}
    \begin{subfigure}[b]{0.24\textwidth}
    \centering
        \includegraphics[width=\columnwidth]{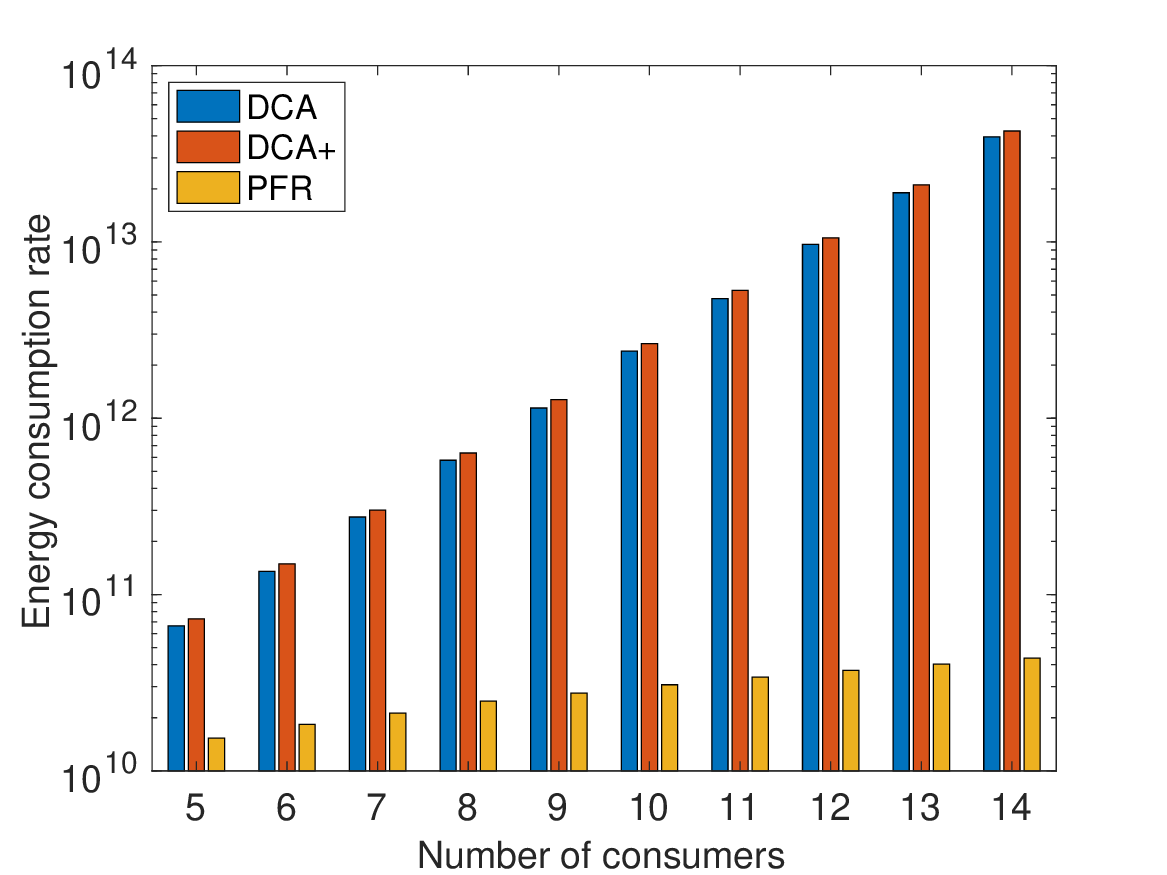}
        \caption{49 nodes.}
        \label{fig::dca3-nrg}
    \end{subfigure}
    \begin{subfigure}[b]{0.24\textwidth}
    \centering
        \includegraphics[width=\columnwidth]{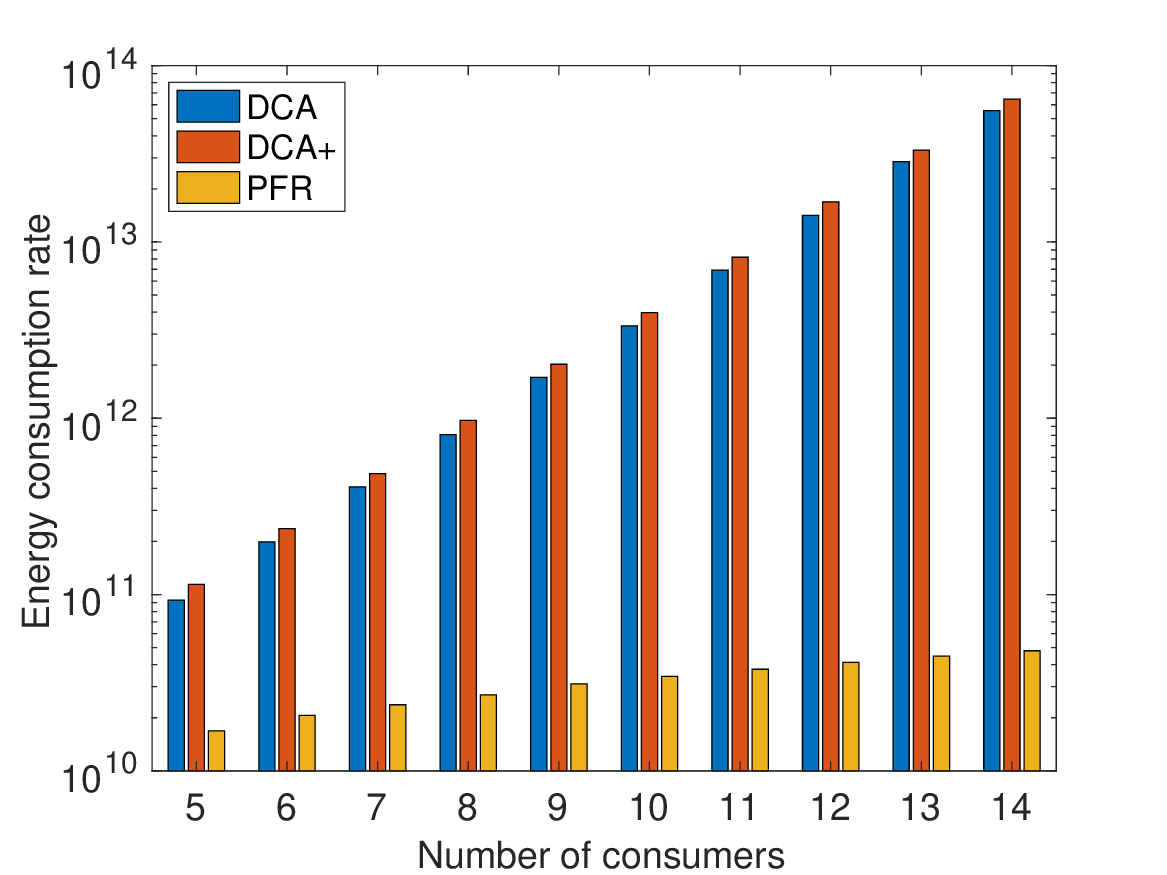}
        \caption{64 nodes.}
        \label{fig::dca4-nrg}
    \end{subfigure}
    
   \begin{subfigure}[b]{0.24\textwidth}
    \centering
        \includegraphics[width=\columnwidth]{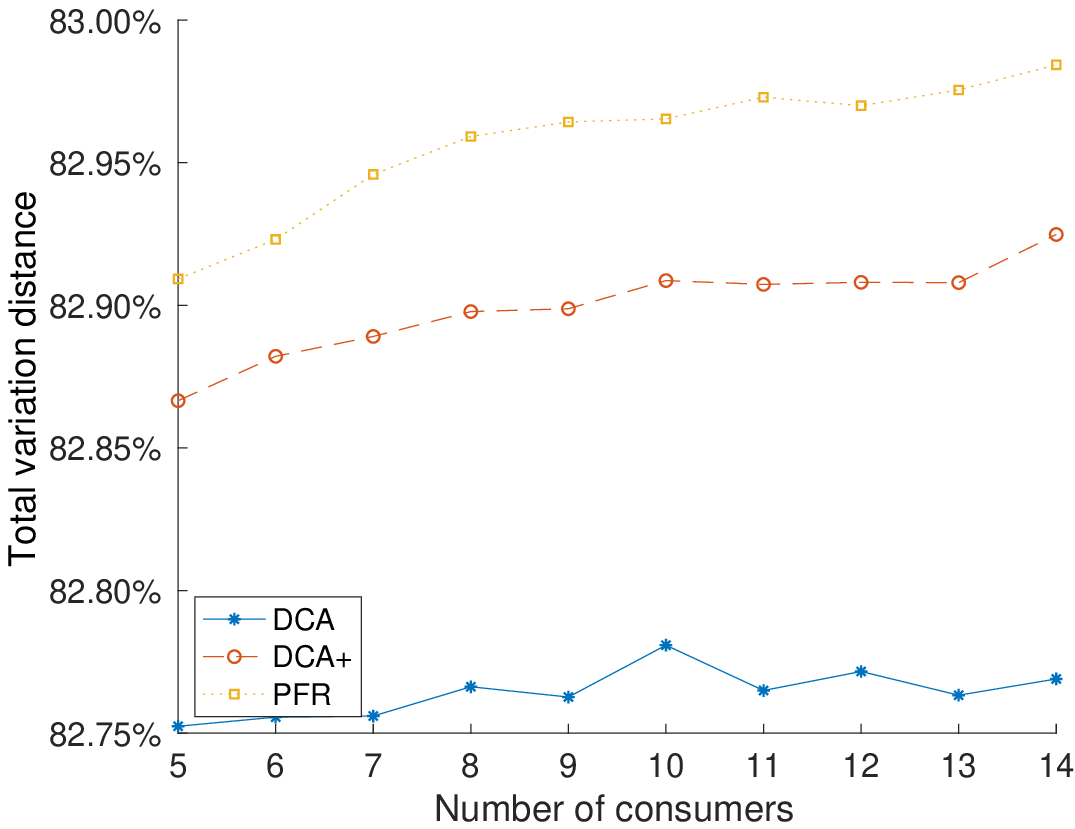}
        \caption{25 nodes.}
        \label{fig::dca1-vd}
    \end{subfigure}
    \label{fig::dca-comp}
    \begin{subfigure}[b]{0.24\textwidth}
    \centering
        \includegraphics[width=\columnwidth]{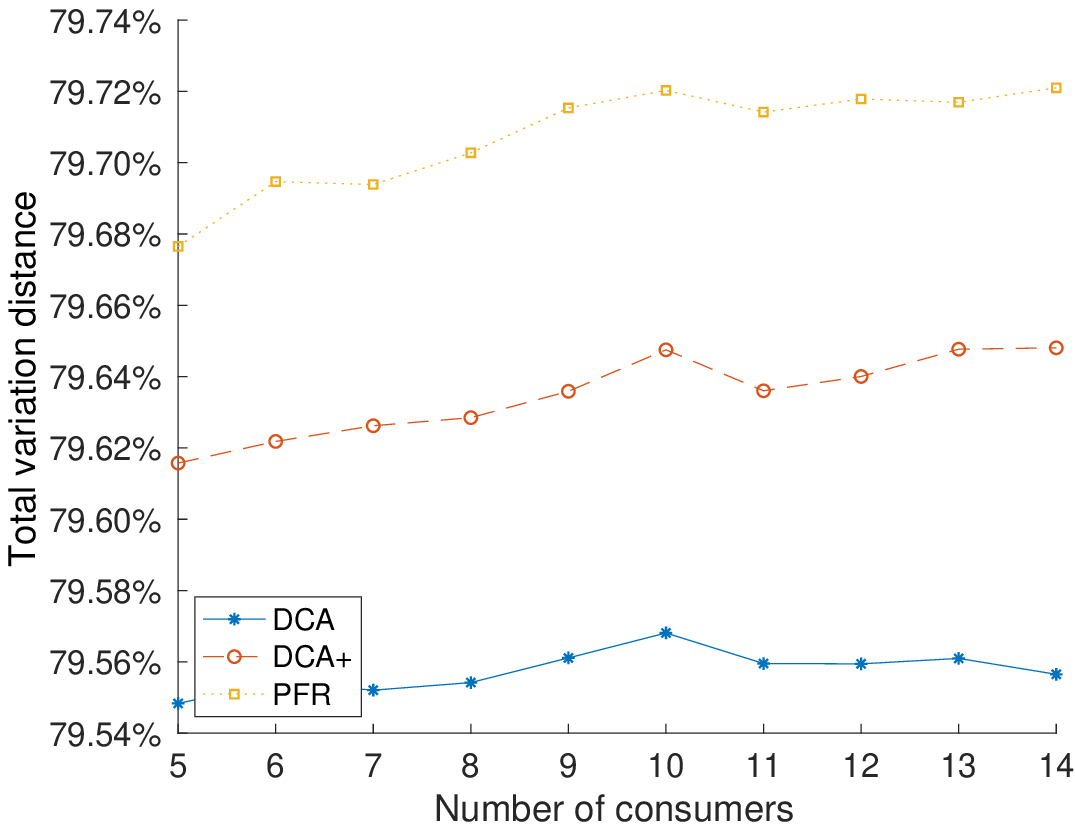}
        \caption{36 nodes.}
        \label{fig::dca2-vd}
    \end{subfigure}
    \begin{subfigure}[b]{0.24\textwidth}
    \centering
        \includegraphics[width=\columnwidth]{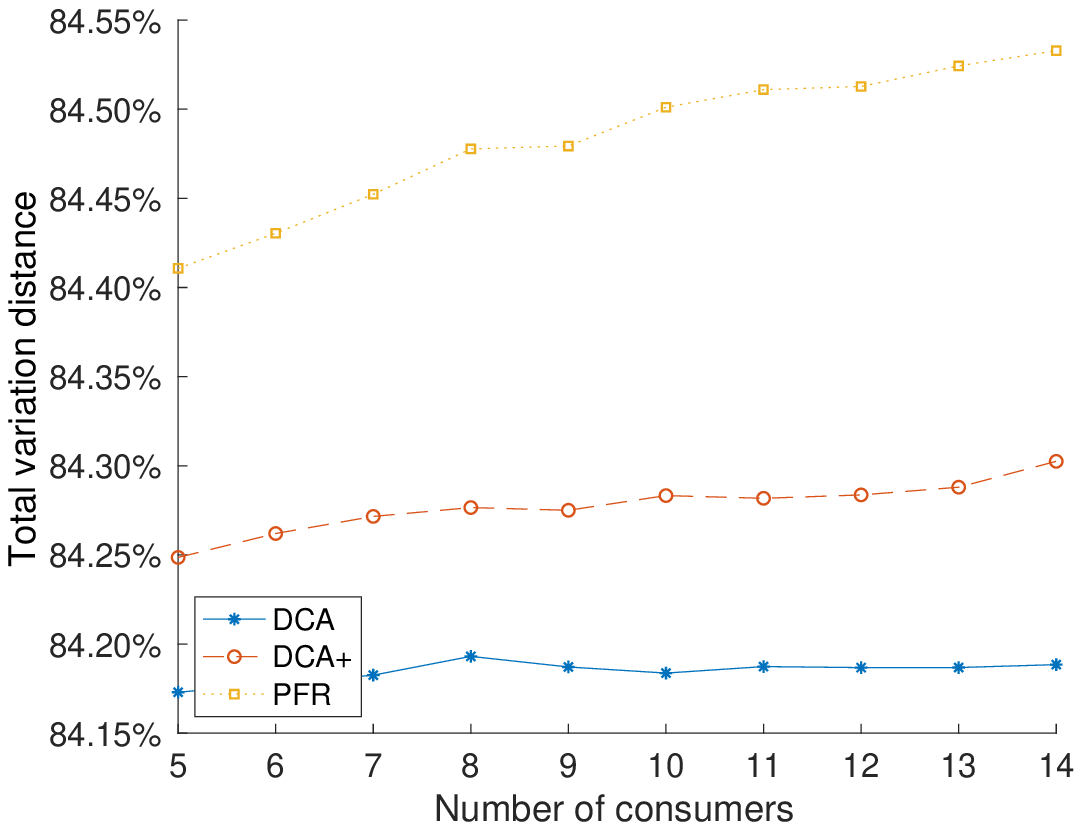}
        \caption{49 nodes.}
        \label{fig::dca3-vd}
    \end{subfigure}
    \begin{subfigure}[b]{0.24\textwidth}
    \centering
        \includegraphics[width=\columnwidth]{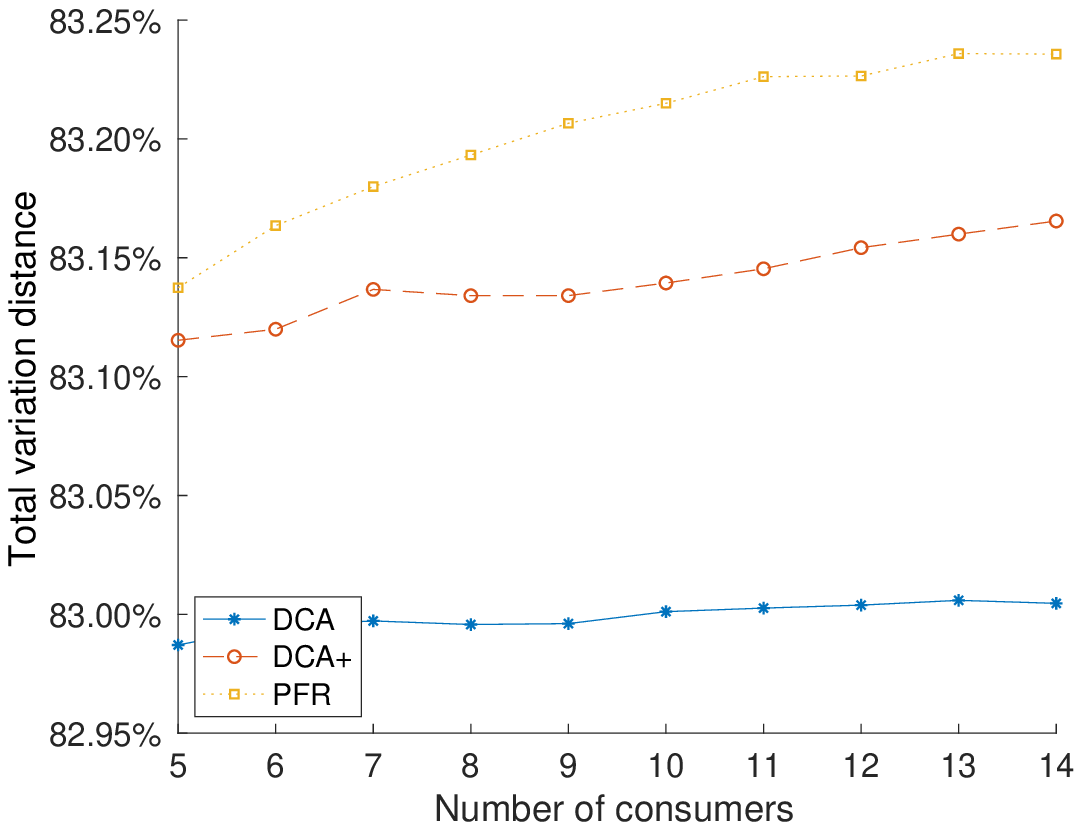}
        \caption{64 nodes.}
        \label{fig::dca4-vd}
    \end{subfigure}
    \caption{Lifetime, energy consumption rate, total variation distance.}
\end{figure*}

\subsection{Simulations and comparison} \label{sec::sims}

We compare via simulations the performance of \texttt{DCA}, \texttt{DCA+} and \texttt{PFR}. The use of simulations instead of field experiments allow us to be more flexible that in the experimental case, and explore larger scale scenaria and different parameter values (we have shown in section \ref{sec::exps} that our simulation model nicely approximates experimental results). The intuition behind the simulations is to maintain the testbed experimental features, but to variate several parameters that were difficult to handle in the testbed experiments. Therefore, we maintain a grid topology for the performance valuation purposes, however, in order to obtain additional results, we variate the size of the topology with increasing numbers of nodes and increasing numbers of consumers. Specifically, we compare the methods on (square) grid network topologies of different sizes, for 25, 36, 49 and 64 numbers of nodes. We focus on the performance of the algorithms on three different metrics: the network lifetime, the energy consumption rate, which is the ratio of the total energy consumed in the network to the network lifetime, and the total variation distance after the simulation run, which is the statistical distance \cite{LevinPeresWilmer2006} between the distribution of node energies in the network at the end of the simulation (i.e., when the first node dies) and the uniform distribution of the total energy among the nodes in the network at the same time. The total variation distance serves as an indicator for the energy balance that each method achieves in the network. We keep the number of proxies close to the small number defined in the experimental parameters, and we use a percentage of cache nodes of $|P| = 0.2|V|$, so as to also maintain the modeling assumption of $|P| \ll |V-P|$. Also, we adopt the same $l^{(h)}, L_{\text{max}}$ values as with the experiments. Additionally, we can simulate higher energy supplies for each node, something that was infeasible in the testbed, as the testbed nodes had a specific fixed amount of energy stored in their batteries. So we set $30-50$ Wh per node and $100-150$ Wh per cache node, data piece generation and consumption rate $r^g_d = r^c_d = 1-8$ $d$/sec. For \texttt{DCA+} and \texttt{PFR} we set $\alpha\tau = 10$ hours. The simulations were conducted in Matlab, and each value in the plots is the result of 500 runs of the same experiment with random (within the aforementioned ranges) initial energy supplies and generation/consumption rates. The statistical analysis of the findings (the median, lower and upper quartiles, outliers of the samples) demonstrate very high concentration around the mean, so in the following figures we only depict average values.

\emph{Network lifetime}. We ran the three methods in networks of increasing number of consumers ($c_d = 5-14$) and for increasing numbers of nodes, and we display the results in Figs.~\ref{fig::dca1}-\ref{fig::dca4}. For purposes of visual convenience, the y-axis of the figures is presented in logarithmic scale. Also in each figure there is a magnified portion of the plot, displaying closely the part of \texttt{DCA} and \texttt{DCA+}. We can see that for increasing number of nodes (with constant percentage of cache nodes in the network), the performance trend follows a similar pattern for all three methods. The conclusion is that even if both \texttt{DCA+} and \texttt{PFR} outperform \texttt{DCA}, the performance gap exhibits significant variation. Specifically, \texttt{DCA+} manages to prolong the network lifetime at the best case for an order of magnitude of $1000$ hours, \texttt{PFR} is prolonging the lifetime for an order of magnitude of $10000$ hours. This remarkably significant improvement comes from the fact that \texttt{PFR} is using only the low-power edge wireless links which are typically operating at $-25$ dBm (without having the privilege of efficient central \texttt{DCA} recomputations), in contrast to  \texttt{DCA+} which is using also local area wireless (and is acquiring updated efficient schedules based on the current energy map of the network), typically at $15$ dBm. Those values translate to power of $3$ $\mu$W and $32$ mW respectively, and therefore exhibit a difference of at least three orders of magnitude. Taking a look at the figures with the performance results, we can empirically validate this phenomenon, as the difference between the performance of \texttt{PFR} and \texttt{DCA+}, interestingly enough, exhibits indeed a difference of 0-3 orders of magnitude.

\emph{Energy consumption rate}. the energy consumption rate is displayed in Figs.~\ref{fig::dca1-nrg}-\ref{fig::dca4-nrg}. Again, the performance trend follows a similar pattern for all three methods, in every instance of the experiment. Specifically, while \texttt{DCA+} manages to reduce the energy consumption rate at the best case within the same order of magnitude, \texttt{PFR} is reducing the energy consumption rate for up to 3 orders of magnitude (for example $10^{10}$ against $10^{13}$ for 14 consumers). This is a result of the local area wireless communications and is quite interesting, as it shows that the advantage offered by \texttt{DCA+}, of recomputing dynamically the best paths as time goes by, is basically nullified by the cost of additional reporting by field nodes towards the central controller. On the contrary, \texttt{PFR}, achieves the significantly lower energy consumption rate ($0-3$ orders of magnitude), thanks to the data distribution path rotation throughout the experiment. Considering together the results related to network lifetime and energy rate consumption, we obtain a very interesting picture. Specifically, a fully distributed algorithm like \texttt{PFR} provides a striking advantage in terms of energy efficiency with respect to a centralized solution. Even though the paths used by \texttt{PFR} might be suboptimal (as opposed to \texttt{DCA+}), the additional cost paid on those paths is way lower than the cost required by \texttt{DCA+} to keep, at the central controller, an updated view of the energy status of each node. Such energy disadvantage of centralized solutions are, to the best of our knowledge, not often highlighted. Together with issues such as additional delays and jitter in communications, they also push towards more decentralized solutions for data management, such as \texttt{PFR}.

\emph{Total variation distance}. We use the total variation distance metric \cite{LevinPeresWilmer2006} as an indicator of the energy balance in the network. The energy balance achieved by a method is an indirect effect of this method on the network's energy distribution. Energy balance can be an important metric of industrial edge deployments which can lead to prevention of uneven load or early disconnections. Let $R, Q$ be two probability distributions defined on sample space ${\cal V}$. The total variation distance $\delta(R, Q)$ between $R$ and $Q$ is $\delta(R, Q) \stackrel{def}{=} \frac{1}{2} \sum_{x \in {\cal V}} |R(x)-Q(x)|$. Equivalently, $\delta(R, Q) = \sum_{x \in {\cal V}: R(x) > Q(x)} (R(x)-Q(x)) = \sum_{x \in {\cal V}: R(x) < Q(x)} (Q(x)-R(x))$. At the end of each experiment, we define the distribution ${\cal E}$ on the sample space of the network nodes ${\cal V}$ as ${\cal E}(u) \stackrel{def}{=} \frac{E_u}{E({\cal V})}$, for any $u \in {\cal V}$, where $E({\cal V}) = \sum_{x \in {\cal V}} E_x$. Furthermore, we denote by ${\cal U}$ the uniform distribution on ${\cal V}$. We will say that the network achieves energy balance of value $\mu$ if and only if $\delta({\cal E}, {\cal U}) \leq \mu$. The total variation distance of the three methods is displayed in Figs.~\ref{fig::dca1-vd}-\ref{fig::dca4-vd}. Since a better energy balance is expressed by lower values of total variation distance, it is apparent that eventually the best balance is achieved by \texttt{DCA}, outperforming \texttt{PFR} by a factor of $0.15 - 0.38\%$. For example, for 49 network nodes, while \texttt{PFR} manages to achieve a total variation distance of 84.5\%, \texttt{DCA+} is achieving 84.3\% and \texttt{DCA} 84.2\% (for 14 consumers). However, note that this is a conclusion regarding only the energy balance, not taking into account the lifetime achieved, which in the case of \texttt{PFR} was significantly higher. Naturally, the longer the lifetime is, the more the energy distribution in the population can variate, thus leading to wider variation distances. Also, since this is the maximal difference, the energy balance is almost the same across the three methods, which is another advantage of \texttt{PFR} in the end. A final point to be made is the co-relation of the results on the energy consumption rate and the total variation distance. As we can see in Figs.~\ref{fig::dca1-nrg}-\ref{fig::dca4-nrg}, \texttt{PFR}, achieves the significantly lower energy consumption rate than \texttt{DCA} and \texttt{DCA+}. However, this improvement comes at a cost: As mentioned earlier (and shown in Figs.~\ref{fig::dca1-vd}-\ref{fig::dca4-vd}), \texttt{PFR} leads to a less balanced energy consumption in the network.

\section{Conclusions} \label{sec::conc}

In this paper, we introduced the problem of distributed data access in multi-hop wireless industrial edge deployments and we prove that it is computationally intractable. We designed a two-step algorithm for solving the computation version of the problem, and we used FIT IoT-LAB open testbed \cite{7389098} for conducting the experimental investigation. We provided two online improvements to recompute paths dynamically based on updated values of energy at nodes. The first one is a centralized method which uses local area wireless communication to renew the data distribution schedules according to the current network energy map and the second one is a distributed method which periodically rotates the available data distribution paths in a proportionally fair manner. We compared the performance of all three methods via simulations for different numbers of network nodes and data consumers, and we showed significant lifetime prolongation and increased energy efficiency when using the proportionally fair rotation method. We believe that these results provide one more dimension supporting the advantage of distributed data management schemes over centralized ones. In addition to the well-known advantage of lower delays and jitter, which are very important in industrial environments, we also show that distributed solutions can be dramatically more energy efficient, as they can rely almost exclusively on cheap local wireless communication technologies, avoiding more costly local area wireless networks, required to quickly communicate with central controllers. Finally, end-users with potential related necessities in actual industrial  environments and shop-floors can adopt our solution per se, and obtain similar approximate results. Also, given that the evaluation results show that the performance of the solution is stable for the specific types of networks presented in this paper, our algorithmic design methodology can be reused for addressing similar future emerging problems in different application scenaria.



\begin{IEEEbiography}[{\includegraphics[width=1in,height=1.25in,clip,keepaspectratio]{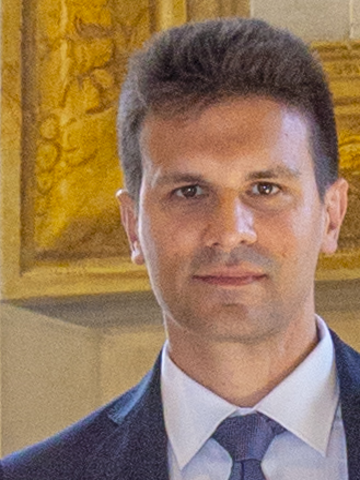}}]{Theofanis P. Raptis}
is a Research Scientist at the National Research Council of Italy. Previously, he was an Associate Researcher at the Computer Technology Institute and Press ``Diophantus'', Greece. He has published in journals, conference proceedings and books more than 60 papers on industrial networks, wirelessly powered networks, internet of things testbeds and platforms. He is serving as Associate Editor for the IEEE Access journal and as Guest Editor for the Elsevier Computer Communications journal. He is regularly involved in international IEEE and ACM sponsored conference and workshop organization committees, in the areas of networks, computing and communications. He obtained his PhD at the University of Patras, Greece.
\end{IEEEbiography}

\begin{IEEEbiography}[{\includegraphics[width=1in,height=1.25in,clip,keepaspectratio]{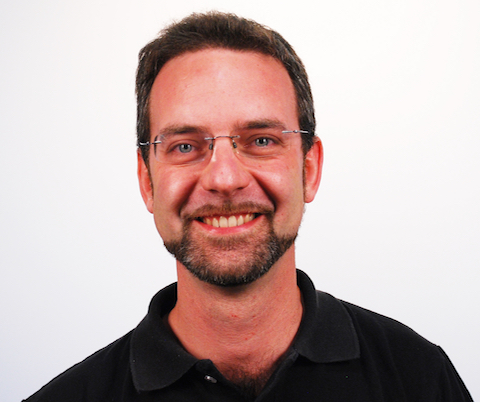}}]{Andrea Passarella}
(PhD 2005) is a Research Director at the Institute for Informatics and Telematics (IIT) of the National Research Council of Italy (CNR). Prior to join IIT he was with the Computer Laboratory of the University of Cambridge, UK. He has published 160+ papers on human-centric data management for self-organising networks, online and mobile social networks, opportunistic, ad hoc and sensor networks. He received four best paper awards, including at IFIP Networking 2011 and IEEE WoWMoM 2013. He was General Co-Chair for IEEE WoWMoM 2019 and workshops co-chair for IEEE INFOCOM 2019. He was the PC co-chair of IEEE WoWMoM 2011, Workshops co-chair of ACM MobiHoc 2015, IEEE PerCom and WoWMoM 2010, and the co-chair of several IEEE and ACM workshops. He is the founding Associate EiC of the new Elsevier journal Online Social Networks and Media (OSNEM). He is co-author of the book ``Online Social Networks: Human Cognitive Constraints in Facebook and Twitter Personal Graphs'' (Elsevier, 2015), and was Guest Co-Editor of the several special issues/sections in ACM and Elsevier Journals and of the book ``Multi-hop Ad hoc Networks: From Theory to Reality'' (2007). He is the chair of the IFIP WG 6.3 ``Performance of Communication Systems''.
\end{IEEEbiography}


\begin{IEEEbiography}[{\includegraphics[width=1in,height=1.25in,clip,keepaspectratio]{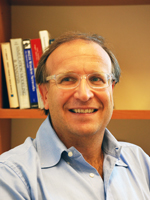}}]{Marco Conti}
is a Research Director and Scientific Counselor, for information and communication technologies, of the Italian National Research Council. He has published in journals and conference proceedings more than 400 scientific papers related to design, modeling, and experimentation of Internet architecture and protocols, pervasive systems and social networks. He is the founding Editor-in-Chief of the Online Social Networks and Media journal, Editor-in-Chief for special issues of the Pervasive and Mobile Computing journal and, from several years, Editor-in-Chief of the Computer Communications journal, all published by Elsevier. He has published the books ``Metropolitan Area Networks (MANs)'' (1997), ``Mobile Ad Hoc Networking'' (2004), ``Mobile Ad hoc networking: the cutting edge technologies'' (2013) and ``Online Social Networks: Human Cognitive Constraints in Facebook and Twitter Personal Graphs'' (2015). He has received several awards, including the Best Paper Award at IFIP TC6 Networking 2011, IEEE ISCC 2012 and IEEE WoWMoM 2013. He served as TPC chair for several major conferences, such as IFIP Networking 2002, IEEE WoWMoM 2005, IEEE PerCom 2006, and ACM MobiHoc 2006, and he was general chair (among many others) for IEEE WoWMoM 2006, IEEE MASS 2007 and IEEE PerCom 2010. He is the founder of successful conference and workshop series, such as IEEE AOC, ACM MobiOpp, and IFIP SustainIT.
\end{IEEEbiography}


\begin{thebibliography}{10}
\providecommand{\url}[1]{#1}
\csname url@samestyle\endcsname
\providecommand{\newblock}{\relax}
\providecommand{\bibinfo}[2]{#2}
\providecommand{\BIBentrySTDinterwordspacing}{\spaceskip=0pt\relax}
\providecommand{\BIBentryALTinterwordstretchfactor}{4}
\providecommand{\BIBentryALTinterwordspacing}{\spaceskip=\fontdimen2\font plus
\BIBentryALTinterwordstretchfactor\fontdimen3\font minus
  \fontdimen4\font\relax}
\providecommand{\BIBforeignlanguage}[2]{{%
\expandafter\ifx\csname l@#1\endcsname\relax
\typeout{** WARNING: IEEEtran.bst: No hyphenation pattern has been}%
\typeout{** loaded for the language `#1'. Using the pattern for}%
\typeout{** the default language instead.}%
\else
\language=\csname l@#1\endcsname
\fi
#2}}
\providecommand{\BIBdecl}{\relax}
\BIBdecl

\bibitem{CONTI20122}
\BIBentryALTinterwordspacing
M.~Conti, S.~K. Das, C.~Bisdikian, M.~Kumar, L.~M. Ni, A.~Passarella,
  G.~Roussos, G.~Tröster, G.~Tsudik, and F.~Zambonelli, ``{Looking ahead in
  pervasive computing: Challenges and opportunities in the era of
  cyber-physical convergence},'' \emph{Pervasive and Mobile Computing}, vol.~8,
  no.~1, pp. 2 -- 21, 2012. [Online]. Available:
  \url{http://www.sciencedirect.com/science/article/pii/S1574119211001271}
\BIBentrySTDinterwordspacing

\bibitem{8012376}
X.~{Xu} and Q.~{Hua}, ``{Industrial Big Data Analysis in Smart Factory: Current
  Status and Research Strategies},'' \emph{IEEE Access}, vol.~5, pp.
  17\,543--17\,551, 2017.

\bibitem{8085101}
J.~{Yan}, Y.~{Meng}, L.~{Lu}, and L.~{Li}, ``{Industrial Big Data in an
  Industry 4.0 Environment: Challenges, Schemes, and Applications for
  Predictive Maintenance},'' \emph{IEEE Access}, vol.~5, pp. 23\,484--23\,491,
  2017.

\bibitem{networld2020}
``{NetWorld2020: Smart Networks in the context of NGI - Strategic Research and
  Innovation Agenda 2021-27},''
  \url{https://www.networld2020.eu/wp-content/uploads/2018/11/networld2020-5gia-sria-version-2.0.pdf},
  accessed on 1 July 2019.

\bibitem{7851047}
L.~{Ascorti}, S.~{Savazzi}, G.~{Soatti}, M.~{Nicoli}, E.~{Sisinni}, and
  S.~{Galimberti}, ``A wireless cloud network platform for industrial process
  automation: Critical data publishing and distributed sensing,'' \emph{IEEE
  Transactions on Instrumentation and Measurement}, vol.~66, no.~4, pp.
  592--603, April 2017.

\bibitem{8259028}
J.~{Fu}, Y.~{Liu}, H.~{Chao}, B.~K. {Bhargava}, and Z.~{Zhang}, ``{Secure Data
  Storage and Searching for Industrial IoT by Integrating Fog Computing and
  Cloud Computing},'' \emph{IEEE Transactions on Industrial Informatics},
  vol.~14, no.~10, pp. 4519--4528, Oct 2018.

\bibitem{8472907}
M.~{Bennis}, M.~{Debbah}, and H.~V. {Poor}, ``Ultrareliable and low-latency
  wireless communication: Tail, risk, and scale,'' \emph{Proceedings of the
  IEEE}, vol. 106, no.~10, pp. 1834--1853, Oct 2018.

\bibitem{5gppp}
``{5GPPP: 5G and the Factories of the Future},'' \url{https://5g-ppp.eu/},
  accessed on 1 July 2019.

\bibitem{Raptis_2018}
\BIBentryALTinterwordspacing
T.~Raptis, A.~Passarella, and M.~Conti, ``{Performance Analysis of
  Latency-Aware Data Management in Industrial IoT Networks},'' \emph{Sensors},
  vol.~18, no.~8, p. 2611, Aug 2018. [Online]. Available:
  \url{http://dx.doi.org/10.3390/s18082611}
\BIBentrySTDinterwordspacing

\bibitem{Lucas_Esta__2018}
\BIBentryALTinterwordspacing
M.~Lucas-Estan, M.~Sepulcre, T.~Raptis, A.~Passarella, and M.~Conti,
  ``{Emerging Trends in Hybrid Wireless Communication and Data Management for
  the Industry 4.0},'' \emph{Electronics}, vol.~7, no.~12, p. 400, Dec 2018.
  [Online]. Available: \url{http://dx.doi.org/10.3390/electronics7120400}
\BIBentrySTDinterwordspacing

\bibitem{7785890}
K.~{Wang}, Y.~{Wang}, Y.~{Sun}, S.~{Guo}, and J.~{Wu}, ``{Green Industrial
  Internet of Things Architecture: An Energy-Efficient Perspective},''
  \emph{IEEE Communications Magazine}, vol.~54, no.~12, pp. 48--54, December
  2016.

\bibitem{4567876}
S.~{Even}, A.~{Itai}, and A.~{Shamir}, ``On the complexity of time table and
  multi-commodity flow problems,'' in \emph{16th Annual Symposium on
  Foundations of Computer Science (sfcs 1975)}, Oct 1975, pp. 184--193.

\bibitem{5677539}
L.~{Mottola} and G.~P. {Picco}, ``{MUSTER: Adaptive Energy-Aware Multisink
  Routing in Wireless Sensor Networks},'' \emph{IEEE Transactions on Mobile
  Computing}, vol.~10, no.~12, pp. 1694--1709, Dec 2011.

\bibitem{6687258}
M.~{Magno}, D.~{Boyle}, D.~{Brunelli}, E.~{Popovici}, and L.~{Benini},
  ``{Ensuring Survivability of Resource-Intensive Sensor Networks Through
  Ultra-Low Power Overlays},'' \emph{IEEE Transactions on Industrial
  Informatics}, vol.~10, no.~2, pp. 946--956, May 2014.

\bibitem{7950207}
B.~{Chen}, L.~{Liu}, Z.~{Zhang}, W.~{Yang}, and H.~{Ma}, ``{BRR-CVR: A
  Collaborative Caching Strategy for Information-Centric Wireless Sensor
  Networks},'' in \emph{2016 12th International Conference on Mobile Ad-Hoc and
  Sensor Networks (MSN)}, Dec 2016, pp. 31--38.

\bibitem{8325491}
M.~J. {Herrmann} and G.~G. {Messier}, ``Cross-layer lifetime optimization for
  practical industrial wireless networks: A petroleum refinery case study,''
  \emph{IEEE Transactions on Industrial Informatics}, vol.~14, no.~8, pp.
  3559--3566, Aug 2018.

\bibitem{6812138}
A.~{Saifullah}, Y.~{Xu}, C.~{Lu}, and Y.~{Chen}, ``End-to-end communication
  delay analysis in industrial wireless networks,'' \emph{IEEE Transactions on
  Computers}, vol.~64, no.~5, pp. 1361--1374, May 2015.

\bibitem{8764545}
T.~P. {Raptis}, A.~{Passarella}, and M.~{Conti}, ``{Data Management in Industry
  4.0: State of the Art and Open Challenges},'' \emph{IEEE Access}, vol.~7, pp.
  97\,052--97\,093, 2019.

\bibitem{8390794}
------, ``{Maximizing industrial IoT network lifetime under latency constraints
  through edge data distribution},'' in \emph{2018 IEEE Industrial
  Cyber-Physical Systems (ICPS)}, May 2018, pp. 708--713.

\bibitem{7389098}
C.~Adjih, E.~Baccelli, E.~Fleury, G.~Harter, N.~Mitton, T.~Noel,
  R.~Pissard-Gibollet, F.~Saint-Marcel, G.~Schreiner, J.~Vandaele, and
  T.~Watteyne, ``{{FIT IoT-LAB: A large scale open experimental IoT
  testbed}},'' in \emph{2015 IEEE 2nd World Forum on Internet of Things
  (WF-IoT)}, Dec 2015, pp. 459--464.

\bibitem{8333734}
Y.~{Luo}, Y.~{Duan}, W.~{Li}, P.~{Pace}, and G.~{Fortino}, ``A novel mobile and
  hierarchical data transmission architecture for smart factories,'' \emph{IEEE
  Transactions on Industrial Informatics}, vol.~14, no.~8, pp. 3534--3546, Aug
  2018.

\bibitem{8291116}
------, ``Workshop networks integration using mobile intelligence in smart
  factories,'' \emph{IEEE Communications Magazine}, vol.~56, no.~2, pp. 68--75,
  Feb 2018.

\bibitem{1435743}
A.~Willig, K.~Matheus, and A.~Wolisz, ``Wireless technology in industrial
  networks,'' \emph{Proceedings of the IEEE}, vol.~93, no.~6, pp. 1130--1151,
  June 2005.

\bibitem{reqs}
``{Network-based communication for Industrie 4.0},'' Publications of Plattform
  Industrie 4.0, \url{www.plattform-i40.de}, 2016, accessed: 14-11-2017.

\bibitem{itg}
A.~M\"{u}ller, ``{Contribution to the discussion session on ''Radio
  communication for Industrie 4.0''},'' {on May, 28th, 2015 at the ITG-expert
  comittee 7.2 radio systems}, 2015.

\bibitem{ANASTASI2009537}
\BIBentryALTinterwordspacing
G.~Anastasi, M.~Conti, M.~D. Francesco, and A.~Passarella, ``Energy
  conservation in wireless sensor networks: A survey,'' \emph{Ad Hoc Networks},
  vol.~7, no.~3, pp. 537 -- 568, 2009. [Online]. Available:
  \url{http://www.sciencedirect.com/science/article/pii/S1570870508000954}
\BIBentrySTDinterwordspacing

\bibitem{Garey}
M.~R. Garey and D.~S. Johnson, \emph{Computers and Intractability: A Guide to
  the Theory of NP-Completeness}.\hskip 1em plus 0.5em minus 0.4em\relax W. H.
  Freeman \& Co., 1979.

\bibitem{1331424}
J.-H. Chang and L.~Tassiulas, ``Maximum lifetime routing in wireless sensor
  networks,'' \emph{IEEE/ACM Transactions on Networking}, vol.~12, no.~4, pp.
  609--619, Aug 2004.

\bibitem{4573259}
V.~Shah-Mansouri, A.~H. Mohsenian-Rad, and V.~W.~S. Wong, ``Lexicographically
  optimal routing for wireless sensor networks with multiple sinks,''
  \emph{IEEE Transactions on Vehicular Technology}, vol.~58, no.~3, pp.
  1490--1500, March 2009.

\bibitem{eppstein}
\BIBentryALTinterwordspacing
D.~Eppstein, ``{Finding the k Shortest Paths},'' \emph{SIAM Journal on
  Computing}, vol.~28, no.~2, pp. 652--673, 1998. [Online]. Available:
  \url{https://doi.org/10.1137/S0097539795290477}
\BIBentrySTDinterwordspacing

\bibitem{yen}
\BIBentryALTinterwordspacing
J.~Y. Yen, ``{Finding the K Shortest Loopless Paths in a Network},''
  \emph{Management Science}, vol.~17, no.~11, pp. 712--716, 1971. [Online].
  Available: \url{http://www.jstor.org/stable/2629312}
\BIBentrySTDinterwordspacing

\bibitem{cc2420}
``{CC2420 datasheet},'' \url{http://www.ti.com/lit/ds/symlink/cc2420.pdf},
  accessed: 01-07-2019.

\bibitem{LM820}
``{LM820 datasheet},''
  \url{https://www.lm-technologies.com/lm_downloads/LM820_DATASHEET.pdf},
  accessed: 01-07-2019.

\bibitem{LevinPeresWilmer2006}
D.~A. Levin, Y.~Peres, and E.~L. Wilmer, \emph{{Markov chains and mixing
  times}}.\hskip 1em plus 0.5em minus 0.4em\relax American Mathematical
  Society, 2006.

\end{thebibliography}
\end{document}